\documentclass[12pt]{siamonline171218}
\usepackage{amsmath}
\usepackage{mathtools}
\usepackage[mathlines]{lineno}
\usepackage{amsfonts}
\usepackage{mathptmx}
\usepackage{times,cite}
\usepackage{svg}
\usepackage{empheq}
\usepackage{microtype}
\usepackage{overpic}
\usepackage{scalerel}
\usepackage[usestackEOL]{stackengine}
\usepackage[export]{adjustbox}
\graphicspath{{figures/}}

\usepackage{lipsum}
\usepackage{epstopdf}
\usepackage{algorithmic}
\ifpdf
  \DeclareGraphicsExtensions{.eps,.pdf,.png,.jpg}
\else
  \DeclareGraphicsExtensions{.eps}
\fi

\usepackage{enumitem}
\setlist[enumerate]{leftmargin=.5in}
\setlist[itemize]{leftmargin=.5in}

\newsiamremark{remark}{Remark}
\newsiamremark{hypothesis}{Hypothesis}
\crefname{hypothesis}{Hypothesis}{Hypotheses}
\newsiamthm{claim}{Claim}

\headers{Generation and motion of interfaces}{P. W. Miller et al.}

\usepackage{xcolor,soul}

\def\XXint#1#2#3{{\setbox0=\hbox{$#1{#2#3}{\int}$}
     \vcenter{\hbox{$#2#3$}}\kern-.5\wd0}}

\newenvironment{sciabstract}{\begin{quote} \bf}{\end{quote}}

\title{Generation and motion of interfaces in a mass-conserving reaction-diffusion system}

\author{Pearson W. Miller,$^{1}$ Daniel Fortunato,$^{2}$, Matteo Novaga$^{3}$,\\ Stanislav Y. Shvartsman$^{1,4,5}$, Cyrill B. Muratov$^{3, 6, \dagger}$ \\
\normalsize{$^{1}$Center for Computational Biology, Flatiron Institute, New York, NY 10010, USA} \\
\normalsize{$^{2}$Center for Computational Mathematics, Flatiron Institute, New York, NY 10010, USA} \\
\normalsize{\parbox[t]{13cm}{\centering $^{3}$Dipartimento di Matematica, Università di Pisa, Largo B. Pontecorvo 5, 56127 Pisa, Italy}} \\
\normalsize{\parbox[t]{13cm}{\centering $^{4}$Department of Molecular Biology, Princeton University, Princeton, NJ 08540, USA}} \\
\normalsize{\parbox[t]{13cm}{\centering $^{5}$Lewis-Sigler Institute for Integrative Genomics, Princeton University, Princeton, NJ 08540, USA}}\\
\normalsize{\parbox[t]{13cm}{\centering $^{6}$Department of Mathematical Sciences, New Jersey Institute of Technology, Newark, NJ 07102, USA}} \\[1.1em]
\normalsize{$^\dagger$To whom the correspondence should be addressed} \\
\normalsize{E-mail: muratov@njit.edu.}}

\makeatletter
\newcommand*{\addFileDependency}[1]{
  \typeout{(#1)}
  \@addtofilelist{#1}
  \IfFileExists{#1}{}{\typeout{No file #1.}}
}
\makeatother

\date{}

\begin{document} 

\maketitle

\begin{sciabstract}
Reaction-diffusion models with nonlocal constraints naturally arise as limiting cases of coupled bulk-surface models of intracellular signalling. In this paper, a minimal, mass-conserving model of cell-polarization on a curved membrane is analyzed in the limit of slow surface diffusion. Using the tools of formal asymptotics and calculus of variations, we study the characteristic wave-pinning behavior of this system on three dynamical timescales. On the short timescale, generation of an interface separating high- and low-concentration domains is established under suitable conditions. Intermediate timescale dynamics is shown to lead to a uniform growth or shrinking of these domains to sizes which are fixed by global parameters. Finally, the long time dynamics reduces to area-preserving geodesic curvature flow that may lead to multi-interface steady state solutions. These results provide a foundation for studying cell polarization and related phenomena in biologically relevant geometries. 
\end{sciabstract}



\section{Introduction}

Reaction-diffusion models are essential tools for understanding the spatial self-organization of chemical patterns inside the cell. Over the past decade, research interest has coalesced around two key features which set intracellular dynamics apart from other classes of models. First, in contrast to traditional models which exist on a single domain, these models are often bulk-surface models, in that they feature distinct diffusion processes within the 3D cytosolic volume of the cell and on the 2D cell membrane coupled by a nonlinear boundary condition \cite{rappel2017mechanisms}. Second, many recent studies have emphasized systems subject to mass-conservation \cite{otsuji2007mass, halatek2018rethinking}. These properties are particularly prevalent in models of cell polarization, a crucial process by which the spatial distribution of proteins within a cell becomes highly localized to a region of the membrane as a result of spontaneous symmetry breaking or an external guiding cue \cite{edelstein2013simple}. Polarization is essential for a great variety of biological phenomena, including guiding developmental outcomes, establishing axes for cell division and guiding locomotion in motile cells, and so has motivated significant interest from the mathematical biology community \cite{edelstein2013simple}.

The attention of applied mathematicians is increasingly turning to understanding how various forms of spatial heterogeneity influence dynamics and steady state behavior. Research that has emphasized the effect of spatially varying kinetic parameters are the most obvious examples of this trend \cite{niethammer2020bulk, ghose2022orientation}. Over the last few years, the question of what role cell geometry plays in guiding polarization has drawn considerable interest. While initial polarization models were reduced to 1D systems, computational advances have enabled numerical studies on 2D and fully 3D domains \cite{diegmiller2018spherical, bialecki2017polarization}. Early steps in addressing this problem have produced simulations that are highly suggestive that localization is closely tied to surface curvature via minimization of interfacial length, but this principle has yet to be demonstrated by formal analysis \cite{cusseddu2019coupled, gessele2020geometric}. Very recently, a novel numerical framework was introduced allowing for efficient simulation of cell polarization models on surfaces of revolution, which in particular motivates this particular study \cite{miller2018geometry}.

In this paper, we perform a formal asymptotic analysis of the surface-bound version of the wave-pinning model first proposed in Ref. \cite{mori2008wave}, which features both mass-conservation and bulk-surface coupling.  This model has attracted considerable interest as a minimal theory for polarization, and has been successfully applied to model systems such as the Rho-GTP pathway \cite{trong2014parameter} or Ezrin polarization in embryonic mouse cells \cite{zhu2020developmental}. The emergence of spontaneous symmetry breaking in this system has been intensely studied, and the dependence of the initial instability of the spatially uniform solution on various parameters is well-characterized \cite{trong2014parameter, madzvamuse2015stability, ratz2014symmetry, paquin2019pattern}.  While previous studies considered the asymptotics of wave-pinning as surface diffusion becomes small, they have generally been concerned with 1D domains or lower orders of perturbation than we discover are needed to capture the full effect of 3D-embedded domain geometry \cite{mori2011asymptotic, cusseddu2019coupled}. Other works which have considered fully 3D domains have limited analysis to linear stability studies \cite{ratz2015turing, gessele2020geometric}. Our research here uses the timescale separation techniques to probe the generation and propagation of interfaces in two-species reaction-diffusion systems to the specific case of the wave-pinning model (for related rigorous studies, see \cite{henry2018multiple, sakamoto2000spatial, chen92}). We note that for single bistable reaction-diffusion equations in the Euclidean setting the generation and propagation of interfaces is by now well understood mathematically \cite{chen92a,bellettini95,alfaro12}. In the context closely related to our problem, results on propagation of interfaces for bistable reaction-diffusion equations on Riemannian manifolds were recently obtained in Refs. \cite{pisante2013allen,pisante2015allen}. 

In this work, we demonstrate that understanding the patterning outcome of polar domains on a surface of arbitrary shape requires a thorough analysis of the long-timescale behavior of the associated mass-conserving reaction-diffusion equation. Working in the limit of slow surface diffusion, we establish a separation of the dynamics into three distinct time scales: first, the initial generation of the interface, then, uniform growth or shrinking of the domains until their areas converge to the steady state values dependent only on the global parameters, and finally, evolution of the interface by area-preserving geodesic curvature flow that results in a finite union of geodesic disks as time goes to infinity. In doing so, we demonstrate that stable steady states with multiple disjoint interfaces exist on biologically plausible domain shapes, something impossible without geometric effects.   

This paper is organized as follows. In Sec. 2, we formulate our model of polarization on a closed surface and define three sub-problems whose limiting behavior should describe the effective dynamics on different asymptotic timescales.  In Sec. 3, we carry out a preliminary analysis of existence and stability of the uniform states and, in particular, identify the parameter regimes in which only nonuniform steady states can be stable. In Secs. 4 through 6, we asymptotically derive the limiting sub-problems as the surface diffusion coefficient tends to zero and examine their dynamical behavior. Finally, in Sec. 7 we perform a number of numerical tests to corroborate the predictions of the asymptotic theory, and in Sec. 8 we make our concluding remarks.  

\section{Model Summary}

Let $\Omega \subset \mathbb{R}^3$ be an open, bounded, connected set with a sufficiently regular boundary $\partial \Omega$. A basic model of cell polarization can be written as (see \cite{diegmiller2018spherical} for more details):
\begin{align}
    \partial_t B &= D_B \nabla^2_{\partial \Omega} B + k_b \left(\beta + \frac{B^\nu}{G^\nu + B^\nu}\right)C - k_d B \label{eq:bulksurf1} & \text{in } \partial \Omega \times (0, T),  \\
    \partial_t C &= D_C \nabla^2_\Omega C, \label{eq:bulksurf2} & \text{in } \Omega \times (0, T), \\
    D_C (\nabla C \cdot \hat{n}) |_{\partial \Omega} &= -k_b \left(\beta + \frac{B^\nu}{G^\nu + B^\nu}\right)C + k_d B. \label{eq:bulksurf3} & \text{in } \partial \Omega \times (0, T), \\
    B(\cdot, 0) &= B_0 &\text{ in } \partial \Omega, \label{eq:bulksurf4} \\
    \ C(\cdot, 0) &= C_0 &\text{ in } \Omega. \label{eq:bulksurf5} 
\end{align}
Here, \eqref{eq:bulksurf1} describes a reaction-diffusion process on the surface $\partial \Omega$ of the surface-bound protein concentration $B$, while \eqref{eq:bulksurf2} gives bulk diffusion of the concentration $C$ of the same protein in the bounded volume $\Omega$, and \eqref{eq:bulksurf3} is the boundary condition coupling the two. The operator $\nabla^2_{\partial \Omega}$ is the Laplace--Beltrami operator on the surface, while $\nabla^2_\Omega$ refers to the standard 3D Laplacian.  While we start with a model where membrane-bound dynamics is already purely two-dimensional, the validity of this class of model as a limiting case of a membrane of finite thickness was demonstrated in Ref. \cite{li2021bulk}.

We are principally interested in the regime where $D_C \gg D_B$, and in this regime it is reasonable to treat the bulk concentration as spatially uniform \cite{diegmiller2018spherical}.  As mass is conserved globally, we define the quantity $C_{\text{tot}}$ as the total number of protein molecules divided by the bulk volume $V$ of $\Omega$, so that the bulk concentration can be expressed as  
\begin{equation}
    C = C_{\text{tot}} - \frac{1}{V} \int_{\partial \Omega} B \, dS.
\end{equation}
Substituting this back into \eqref{eq:bulksurf1} yields a non-local equation for surface-bound species
\begin{align} 
    \partial_t B = D_B \nabla^2_{\partial \Omega} B - k_d B + k_b \left( \beta + \frac{B^\nu}{G^\nu + B^\nu} \right) \left( C_\text{tot} - \frac{1}{V} \int_{\partial \Omega} B \, dS \right) & \text{ on } \partial \Omega \times (0, T).
\end{align}
This equation is rendered dimensionless by a rescaling $B(x, t) \rightarrow u(\tilde{x}, \tilde{t})$, with $u = k_d B / (k_b C_{\text{tot}})$, $\tilde{t} = k_d t$ and $\tilde{x} = x/\sqrt{A} \in \partial \tilde \Omega$, where $A = |\partial \Omega|$ is the surface area of $\partial \Omega$, and $\tilde \Omega = \Omega / \sqrt{A}$ is a rescaling of $\Omega$ that ensures that the rescaled domain $\tilde \Omega$ has boundary of unit area. 
Dropping the tildes and the subscript $\partial \Omega$ for simplicity of notation from now on, we arrive at the dimensionless form
\begin{align}\label{eq:surfdiff}
    \partial_t u &= \delta^2 \nabla^2 u - u + f(u) \left(1 - \alpha U  \right) & \text{ on } \partial \Omega \times (0, T), \\
 U(t) &= \int_{\partial \Omega} u(x,t) dS & \text{for } t \in (0, T), \\
 u(x, 0) &= g(x) & \text{for } x \in \partial\Omega, \label{eq:surfic}
\end{align}
with the dimensionless parameters\footnote{We use a slightly different convention from \cite{diegmiller2018spherical}, in which the definitions of $\alpha$ and $\delta$ differ from the present ones by constant factors.}
\begin{align}
\delta = \sqrt{D_B \over k_d A}, \qquad \alpha = {k_b A \over k_d V}, \qquad \gamma = {k_d G \over k_b C_{\text{tot}}},
\end{align} 
and where we defined $f(u) = \beta + \frac{u^\nu}{\gamma^\nu + u^\nu}$ for further notational convenience. 
Together with the already dimensionless parameters $\beta$ and $\nu$, the parameters $\alpha$, $\gamma$ and $\delta$ define the parameter space for our problem. As was already noted, we have $|\partial \Omega| = 1$ now. 

The long timescale dynamics of this problem for a purely 1D model was previously studied via asymptotic expansions in \cite{mori2011asymptotic}, and later on a disc in \cite{cusseddu2019coupled}. In Ref. \cite{diegmiller2018spherical}, exact solutions were constructed for the special case of a spherical domain and $\nu = \infty$.  Here we confine ourselves to the specific case of $\nu = 2$ for the sake of analytical tractability, but treat general spatial domains. A few useful remarks on this system can be made based upon existing results. First, \eqref{eq:bulksurf1}--\eqref{eq:bulksurf5} represents a special case of a more general class described in Ref. \cite{sharma2016global}, and per the results therein, there exists a unique classical solution $(B, C)$ such that the functions $B$ and $C$ are smooth and uniformly bounded. Further, Theorem $2.6$ of Ref. \cite{hausberg2018well} demonstrates that our shadow system \eqref{eq:surfdiff}--\eqref{eq:surfic} likewise has a unique weak solution, and that solution is the limit of the solutions to \eqref{eq:bulksurf1}--\eqref{eq:bulksurf5} as $D_B \rightarrow \infty$, justifying the use of this reduced approach.  

In the asymptotic analysis described in the following sections, we find it convenient to frame our problem as three equivalent problems parametrized by $\delta \ll 1$:
\begin{align}
    (P_0^\delta)& \qquad    
     \partial_t u_0^{\delta} = \delta^2 \nabla^2 u_0^{\delta} - u_0^{\delta} + f(u_0^{\delta}) \left(1 - \alpha \int_{\partial \Omega} u_0^{\delta} dS \right), \qquad 
     u_0^{\delta}(x, 0) = g_0^\delta(x), \notag \\
(P_1^\delta)& \qquad    
     \partial_t u_1^{\delta} = \delta \nabla^2 u_1^{\delta} + \delta^{-1}\left[ - u_1^{\delta} + f(u_1^{\delta}) \left(1 - \alpha \int_{\partial \Omega} u_1^{\delta} dS \right) \right], \qquad
     u_1^{\delta}(x, 0) =g_1^\delta(x), \notag 
 \\
(P_2^\delta)& \qquad    
     \partial_t u_2^{\delta} =  \nabla^2 u_2^{\delta} + \delta^{-2} \left[ - u_2^{\delta} + f(u_2^{\delta}) \left(1 - \alpha \int_{\partial \Omega} u_2^{\delta} dS \right) \right],  \qquad
     u_2^{\delta}(x, 0) =g_2^\delta(x). \notag
\end{align}
Each problem describes the dynamics at the timescale of a different order in $\delta$: $(P_0^\delta)$ is the problem on the original $O(1)$ timescale, $(P_1^\delta)$ is the problem on the  $O(\delta^{-1})$ long timescale and $(P_2^\delta)$ is the problem on the  $O(\delta^{-2})$ longer timescale. Crucially, we will show that as $\delta \to 0$ the initial condition of the second and third problem may be taken to be the infinite time limit of the solution at the previous stage, thus allowing to connect the solutions at different timescales. In our analysis below, we will formally obtain the limit behavior of each of these problems when $\delta \rightarrow 0$.

\section{Preliminaries} 
\label{sec:pre}

\begin{figure}\label{f_uniform1}

    \centering
    \includegraphics[scale = 0.5]{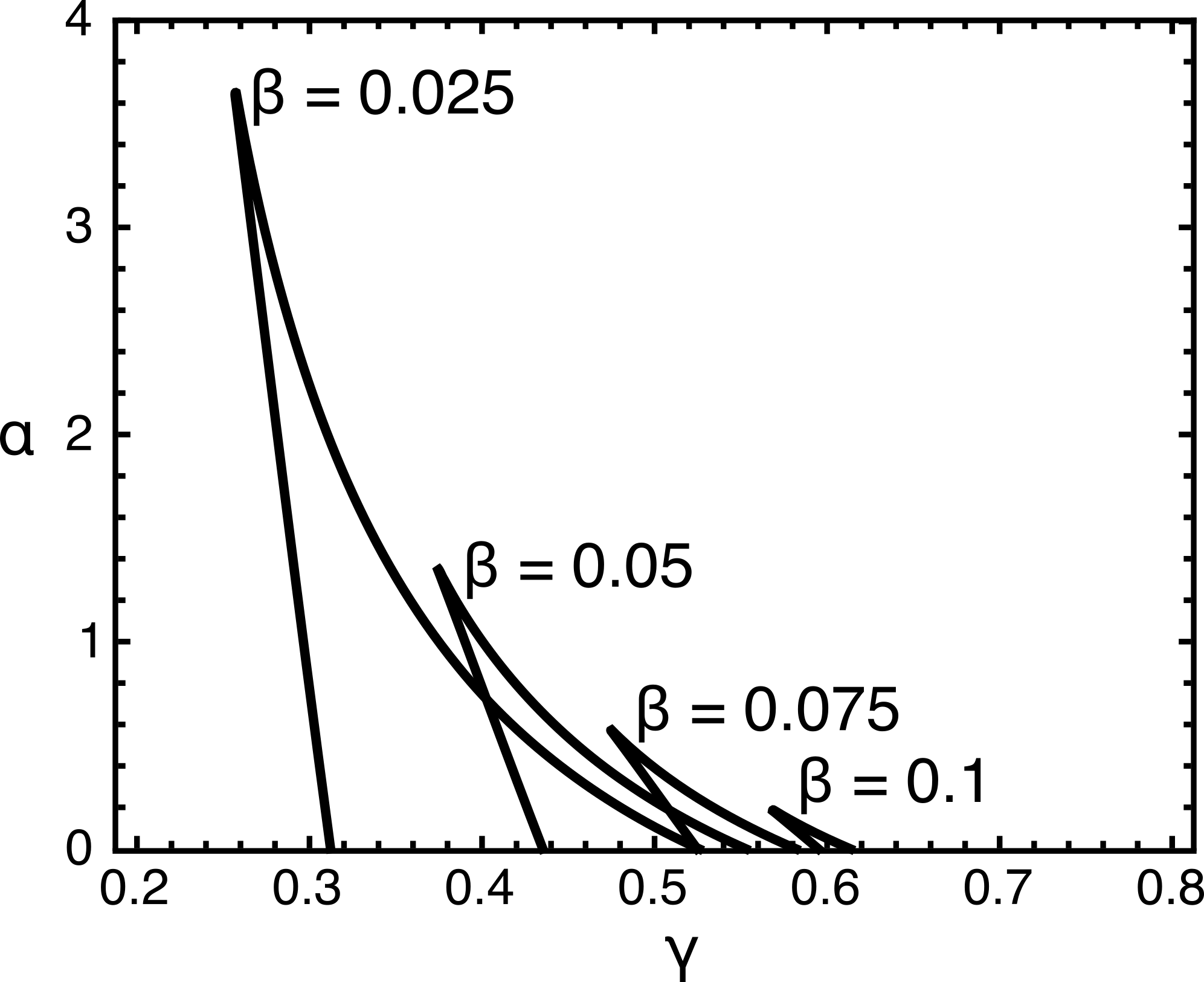}
    \caption{Boundaries between the bistable and monostable regimes plotted for different values of $\beta$. Within the triangular-shaped regions, three positive roots of \eqref{eq:cubic} exist, while outside there is only one.}
\end{figure}

In this section we introduce some basic facts about our system, and establish some notation. We begin by examining the behavior of the spatially uniform steady states of \eqref{eq:surfdiff}. These states satisfy 
\begin{equation}\label{ode}
    -u + f(u)(1 - \alpha  u) = 0,
\end{equation}
and with our choice of the nonlinearity (recall that $\nu = 2$) this reduces to a cubic equation 
\begin{align}
\label{eq:cubic}
    -u(u^2 + \gamma^2) + (1 -\alpha u)(u^2(1 + \beta) + \beta \gamma^2) = 0. 
\end{align} 
For $\alpha, \beta, \gamma > 0$, \eqref{eq:cubic} will have at least one real positive root $u_0$, with two additional positive roots $u_\pm$ emerging when its discriminant 
\begin{align}
\begin{split}
\Delta(\alpha, \beta, \gamma) = \gamma ^4 (1 & -4 \beta 
   (\alpha  \beta +\alpha +1) (2 \beta  (\alpha  \beta +\alpha +1)-5)) \\ 
   & -4 \gamma ^6 (\alpha  \beta +1)^3 (\alpha  \beta +\alpha +1)-4 \beta  (\beta
   +1)^3 \gamma ^2 > 0.    
\end{split}
\end{align} 
We adopt the convention $u_- < u_0 < u_+$, with $u_{\pm,0}$ denoting any one of these roots for shorthand. Setting $\Delta = 0$ allows one to derive the conditions for the existence of three roots. We find the requirements for three real roots to be
\begin{align}
    0 &< \beta < \frac{1}{8},\\
    0 &< \alpha <  \frac{1-8\beta}{8\beta (1 + \beta)},\\
    \gamma_-(\alpha, \beta) &< \gamma < \gamma_+(\alpha, \beta),
\end{align}
where  
\begin{equation}\label{eqn:rootbound}
    \begin{split}
    \gamma_\pm(\alpha, \beta) =  \sqrt{ \frac{1-4 \beta  (\alpha  \beta +\alpha +1) (2 \beta  (\alpha  \beta +\alpha +1)-5) \pm \sqrt{(1 -8 \beta (\alpha \beta + \alpha + 1))^3}}{8 (\alpha  \beta +1)^3 (\alpha  \beta +\alpha +1)}}.
\end{split}
\end{equation}
Slices of the boundary between parameter space regions are plotted in the $\gamma-\alpha$ plane in Fig. \ref{f_uniform1} for various values of $\beta$. 

We next examine the linear stability of the uniformly stable states in the usual manner, examining the behavior of perturbations of the form $u_{\pm,0}(x, t) = u_{\pm, 0} + \epsilon\eta_{\pm, 0}(x, t)$ for $\epsilon \ll 1$. 
Denoting the orthonormal eigenfunctions of the Laplace-Beltrami operator as $v_k(x)$ such that $-\nabla^2 v_k(x) = \lambda_k v_k(x)$ for $k = 0, 1, ...$, where $\lambda_k$ are the corresponding eigenvalues, we write $\eta_{\pm, 0}(x,t) = \sum_{k=0}^\infty a_k^{\pm, 0} e^{\sigma_k^{\pm,0} t} v_k(x)$. Note that for a compact surface, the eigenvalues $\lambda_k$ are real with $0 = \lambda_0 < \lambda_1 \le \lambda_2 \le ...$.  

The linearized equation about the root $u^{0, \pm}$ takes the form   
\begin{align}\label{eqn:linearized}
    \mathcal{L}\eta_{\pm, 0}(x, t) &= \left[\partial_t - \delta^2 \nabla^2 + 1 - f'(u_{\pm, 0})(1-\alpha u_{\pm, 0})\right] \eta_{\pm, 0}(x, t)  + \alpha f(u_{\pm, 0}) \int_{\partial \Omega} \eta_{\pm, 0}(x, t) dS \nonumber \\ 
    &= \sum_{k=0}^\infty \left(\sigma_k^{\pm, 0} + \delta^2 \lambda_k + 1 - f'(u_{\pm, 0})(1-\alpha u_{\pm, 0}) + \alpha f(u_{\pm, 0}) \delta_{k,0} \right) a_k^{\pm, 0} e^{\sigma_k^{\pm, 0} t} v_k(x) =  0 ,
\end{align}
where $\delta_{k,0}$ is the Kronecker delta symbol, and we took into account that $v_0 = 1$ and $\int_{\partial \Omega} v_k dS = 0$ for all $k > 0$.  Then the following dispersion relation holds:
\begin{equation}
    \sigma_k^{\pm, 0} = -\delta^2 \lambda_k - 1 +  f'(u_{\pm, 0})(1-\alpha u_{\pm, 0}) - \alpha f(u_{\pm, 0}) \delta_{k,0}.
\end{equation}
In the three-root parameter region, $\sigma_0^{\pm} < 0$ and $\sigma_0^0 > 0$, while outside this region, $\sigma_0^0 < 0$, as can be verified  numerically. Furthermore, one can see that if $\delta$ is sufficiently large then $\sigma_k^{\pm,0} < 0$ for all $k > 0$. So, as expected, for sufficiently large diffusion the stability of each fixed point can be determined entirely by its stability with respect to uniform perturbation corresponding to $k = 0$.

Conversely, for sufficiently small values of $\delta$ the steady states $u_{\pm,0}$ may lose stability with respect to non-uniform perturbations corresponding to $k > 0$. In this case the fastest growing mode corresponds to $k = 1$, and for $f'(u_{\pm, 0})(1-\alpha u_{\pm, 0}) - 1 > 0$ one can define
\begin{equation}
    \delta_{0, \pm} = \sqrt{\frac{f'(u_{\pm, 0})(1-\alpha u_{\pm, 0}) - 1}{\lambda_1}}. 
\end{equation}
Then for  $\delta < \delta_{0, \pm}$ it follows that $\sigma_1^{0,\pm} > 0$. 

\begin{figure}\label{fig:stab}
    \centering
    \includegraphics[scale = 0.5]{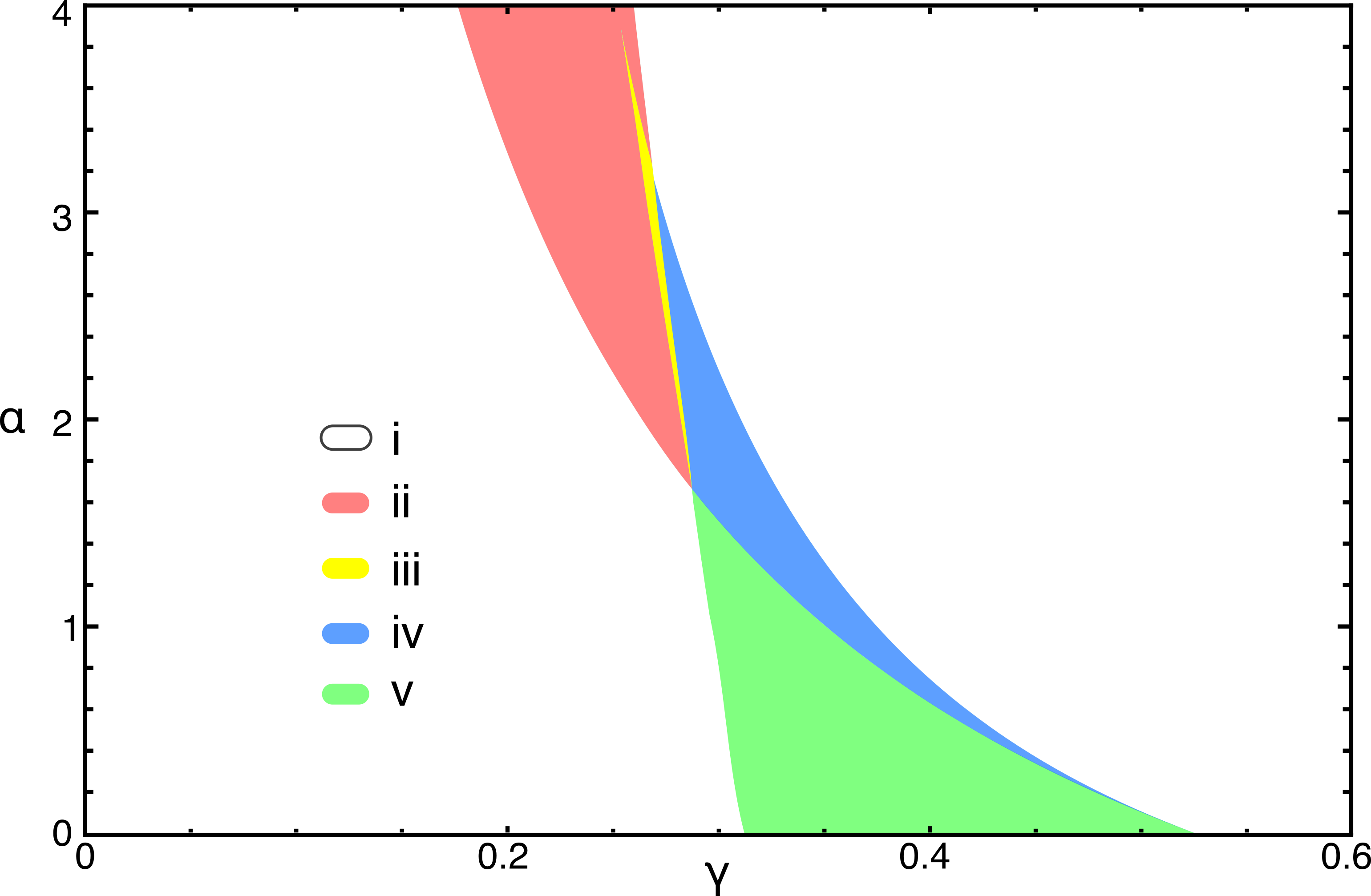}
    \caption{Regions of the parameter space at $\beta = 0.025$ colored according to the stability of the uniform steady state in the limit $\delta \rightarrow 0$. Regions $i$ and $ii$ denote where there is a unique uniform steady state $u_0$. In the former, $\sigma_k^0 < 0$ for all $k$, while in the latter $\sigma_k^0 > 0$ for some $k > 0$.  In $iii$, $iv$ and $v$, three uniform steady states $u_- < u_0 < u_+$ exist, with $\sigma_0^{\pm} < 0$ and $\sigma_0^0 > 0$. In $iv$, $\sigma_k^+> 0$ for some $k > 0$ while $\sigma^-_k < 0$ for all $k$, and in $iii$, $\sigma_k^\pm > 0$ for some $k > 0$. Finally, in $v$, $\sigma^\pm_k < 0$ for all $k$. A final region exists on the boundary of $v$ and $i$ where $u_-$ becomes destabilized but $u_+$ remains stable, but this region is too narrow to reasonably visualize on this plot. }
\end{figure}

As our subsequent analysis emphasizes the limit where $\delta \rightarrow 0$,  it is useful to examine the boundaries of the region where $f'(u_{\pm, 0})(1-\alpha u_{\pm, 0}) = 1$ for each fixed point. 
Let $b_1(\alpha, \gamma) < b_2(\alpha, \gamma)$ be the two real roots of the depressed quartic 
 \begin{equation}
     b^4 + 2\gamma^2(1 +  \alpha)b^2 - 2 \gamma^2 b + \gamma^4 = 0,
 \end{equation}
 which arises from rearranging $f'(b)(1-\alpha b) = 1$.  A uniform fixed point at $u_{\pm,0}$ is stable in the limit $\delta \rightarrow 0$ provided $b_1(\alpha, \beta, \gamma) < u_{\pm,0}(\alpha, \beta, \gamma) < b_2(\alpha, \beta,\gamma)$. We plot a slice of the parameter space at $\beta = 0.025$ in Fig. \ref{fig:stab} in order to illustrate how the parameter space is partitioned. One can see that $\sigma_1^0 > 0$ for all points within the three root region, whereas $\sigma_1^\pm > 0$ in distinct but overlapping regions. Importantly, there is a large parameter region in which no uniform state is linearly stable and, therefore, pattern formation must  occur from a generic initial condition. Note that for finite $\delta$ the diffusion exerts a stabilizing effect and shrinks the regions with positive growth rates, but even for a relatively large value of $\delta$ (e.g. $\delta \approx 0.1$) the stability regions look nearly indistinguishable from those in Fig. \ref{fig:stab}. As a final remark, previous examinations of wave-pinning models have observed that even in the absence of linear instability, such systems tend to be easily destabilized by small finite perturbations - it is hypothesized that many real systems exist entirely within the stable regime but nonetheless polarize due to large scale fluctuations \cite{mori2011asymptotic}.

\section{Generation of Interface}
\label{sec:gen}

At the zeroth order in $\delta$, the diffusion term in $(P_0^\delta)$ may be dropped to the leading order to obtain a closed system of equations for $u_0 = \lim_{\delta \to 0} u_0^\delta$:
\begin{equation}
\label{eqn:zero}
(P_0^0) \qquad
\begin{aligned}
\partial_t u_0 &= - u_0 + f(u_0) (1 - \alpha U_0), \\ 
u_0(x, 0) &= g(x), \\
U_0(t)  &=  \int_{\partial \Omega} u_0(x, t) dS,  
\end{aligned}
\end{equation}
provided $g = \lim_{\delta \to 0} g_0^\delta$ is smooth and the limit is uniform. All spatial coupling now occurs via the parameter $U_0$. In essence, this problem reduces to an infinite system of ODEs connected via a mean-field term. Our goal in this section is to demonstrate that at almost every point $x\in \partial \Omega$, the value of $u(x, t)$ is expected to converge to one of two specific steady state values as $t \to \infty$ for generic choices of $g$. This problem has been considered before in previous studies of wave-pinning, and conventionally the local bistability of (\ref{eqn:zero}) for fixed $U_0$ was used to justify the assumption that $u_0$ indeed converges \cite{mori2011asymptotic, cusseddu2019coupled}. Nevertheless, for variable $U_0$, an additional non-trivial argument is necessary to make such a conclusion. Here, we demonstrate that problem $(P_0^0)$ is in fact a gradient flow generated by the following energy functional:
\begin{equation}
\label{eq:H}
    H[u] = \int_{\partial \Omega} \Phi(u) dS  + \frac{\alpha}{2} \left[\int_{\partial \Omega} u dS \right]^2,
\end{equation}
where
\begin{equation}
    \Phi(u) = \frac{u(1 + \beta)(u - 2(1 + \beta)) + \gamma^2 \log\left(u^2(1+\beta) + \beta \gamma^2) \right)}{2(1+\beta)^2}.
\end{equation}
 This function is bounded from below for all $\beta > 0$, and the second term in \eqref{eq:H} is nonnegative, so we have that $H[u]$ is bounded from below. Taking the variational derivative  of $H$ yields 
\begin{equation}
\partial_t u_0 = -f(u_0) \frac{\delta H}{\delta u}, \qquad \frac{\delta H}{\delta u} = -1 + \frac{u(u^2 + \gamma^2)}{u^2(1 + \beta) + \beta \gamma^2} + \alpha \int_{\partial \Omega} u dS,    
\end{equation}
which is a gradient flow generated by the energy functional $H[u]$ with the mobility $f(u)$. As $f(u_0) > 0$ uniformly, we have
\begin{equation}
\label{eq:dHdt}
    \frac{d H(u_0(\cdot, t))}{dt} = - \int_{\partial \Omega} (f(u_0))^{-1} |\partial_t u_0|^2  dS \le 0.
\end{equation}

\noindent Here equality only occurs when $u_0$ is one of the steady state solutions to (\ref{eqn:zero}), which due to the dissipative nature of the dynamics are the only elements of the attractor of the long-time dynamics of the solutions of $(P_0^0)$ \cite{hale}. Away from the steady states, $H[u_0(\cdot, t)]$ is both decreasing in time and bounded below, so as $t \rightarrow \infty$, $H(u_0(\cdot, t)) \rightarrow H^\infty$. We note that existence of solutions for problem $(P_0^0)$ with positive bounded initial data follows from the standard theory of ODEs in Banach spaces, yielding a unique solution $u_0 \in C^\infty([0, \infty); L^2(\partial \Omega))$ \cite{cartan}. In particular, this gives $U_0 \in C^\infty([0, \infty))$. Therefore, by the standard existence theory for ODEs we have that $u(x, \cdot) \in C^\infty([0, \infty))$ for each $x \in \partial \Omega$. In fact, it is not difficult to see that the solutions of $(P_0^0)$ remain positive and bounded independently of $t$, as they should. Thus, the arguments leading to \eqref{eq:dHdt} are justified. Furthermore, the boundedness of $u_0$ yields boundedness of $\partial_t^2 u_0$ as well, which together with \eqref{eq:dHdt} implies that $\partial_t u_0(\cdot, t) \rightarrow 0$ in $L^2(\partial \Omega)$ as $t \to \infty$. 

\begin{figure}\label{fig:h}
    \centering
    \includegraphics[scale = 0.5]{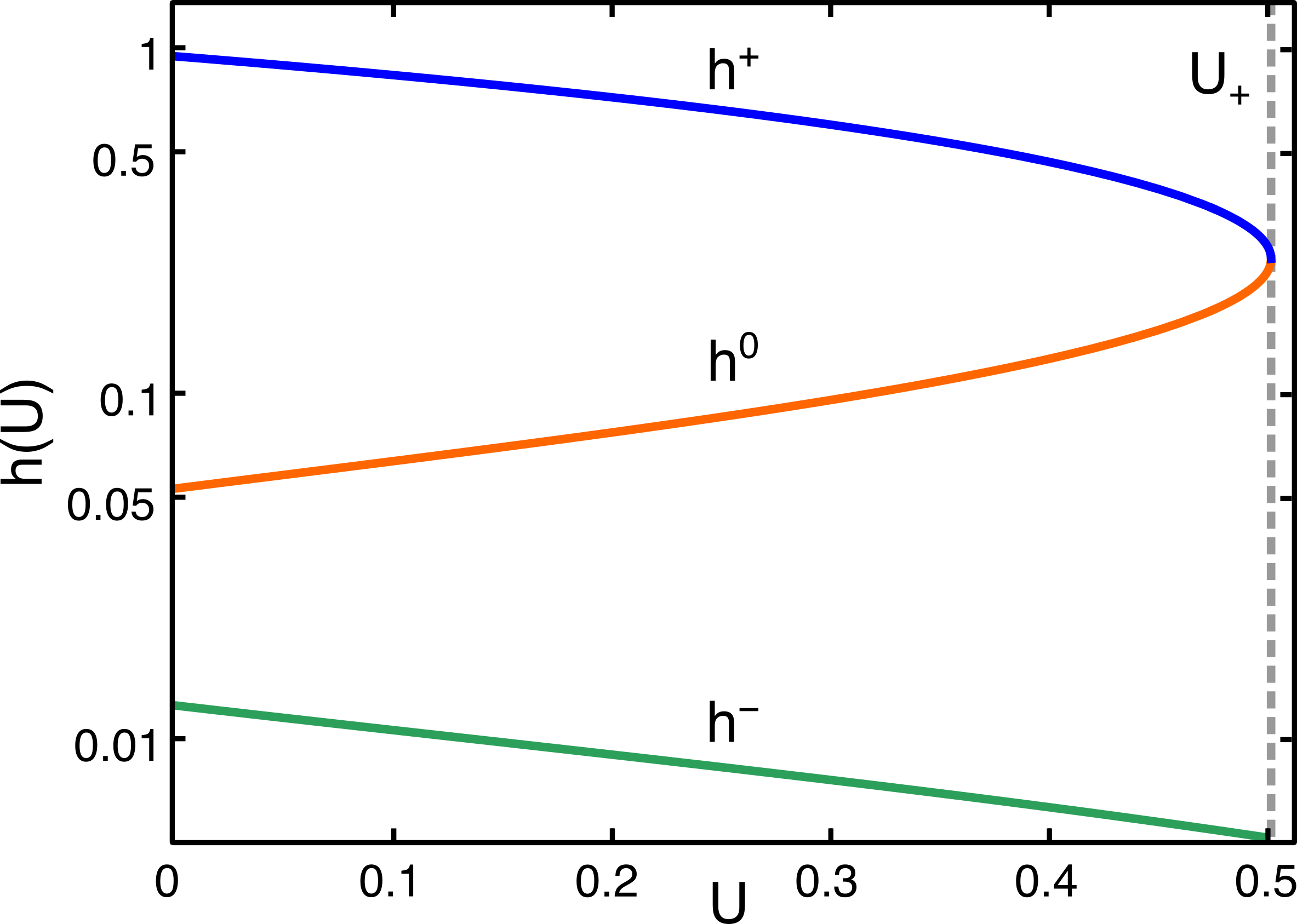}
    \caption{Three solutions $h^+$, $h^0$, and $h^-$ of \eqref{eq:Phiprime} as functions of  $U$.  Here $U_+$ denotes the location of the saddle-node bifurcation of $h^+$ and $h^0$. Parameters chosen for this plot are $\alpha = 1$, $\beta = 0.01$ and $\gamma = 0.25$.} 
\end{figure}

Consider a sequence of $t_n > 0$ such that $t_n \to \infty$ as $n \to \infty$. Up to an extraction of a subsequence (not relabeled), we then have $\partial_t u_0(\cdot, t_n) \to 0$ a.e. in $\partial \Omega$. Since $U_0(t_n)$ is bounded, upon a further extraction of a subsequence we have $U_0(t_n) \to U_0^\infty$ as $n \to \infty$. Therefore, for a.e. $x \in \partial \Omega$ we have that $u(x, t_n)$ converges to one of the roots of the equation
\begin{equation}
\label{eq:Phiprime}
-\Phi'(h) = 1 - \frac{h(h^2 + \gamma^2)}{h^2 ( 1 + \beta) + \beta \gamma^2} = \alpha U
\end{equation}
for $U = U_0^\infty$. This equation has at most three real roots, which we label $h^-(U) < h^0(U) < h^+(U)$, and plot their dependence on $U$ in Fig. \ref{fig:h}.  The conditions for all three roots to be real are $0 < \beta < 1/8$ and $U_-(\alpha, \beta, \gamma) < U < U_+(\alpha, \beta, \gamma)$, where $U_\pm(\alpha, \beta, \gamma)$ correspond to saddle node bifurcations. We calculate the explicit dependence of these on the kinetic parameters as
\begin{align}\label{monostable}
U_\pm(\alpha, \beta, \gamma) &= \alpha^{-1} \left[ 1 - \frac{\left(3 \pm \sqrt{1 - 8 \beta}\right) \sqrt{1 \pm \sqrt{1 - 8 \beta} - 2 \beta)} }{\sqrt{2}(1 \pm \sqrt{1 - 8 \beta} )(1 + \beta)^{3/2}} \gamma \right].
\end{align}
Notably, the form of $\Phi'(h)$ implies that $\frac{dh^\pm(U)}{dU} < 0$. Therefore, the roots $u_0 = h^\pm (U_0^\infty)$ represent the two linearly stable equilibria of \eqref{eqn:zero}, while $u_0 = h^0(U_0^\infty)$ represents an unstable equilibrium for $U_0 = U_0^\infty$ fixed. Thus, generically we would expect that for almost every $x \in \partial \Omega$ we have
\begin{equation} \label{eqn::v_0} 
\lim_{n \to \infty} u_0(x, t_n) = \begin{cases}
h^+(U_0^\infty)  & \text{if} \  x \in \partial \Omega_0^+(U_0^\infty), \\
h^-(U_0^\infty)  & \text{if} \ x \in \partial \Omega_0^-(U_0^\infty),
\end{cases}
\end{equation}
for some $\partial \Omega_0^\pm \subset \partial \Omega$ (possibly depending on the choice of the sequence) such that $|\partial \Omega \backslash (\partial \Omega_0^+ \cup \partial \Omega_0^-) | = 0$. 

From the definitions of $\partial \Omega_0^+$,  $U_0^\infty = \lim_{n \to \infty} \int_{\partial \Omega} u_0(x, t_n) dS $, and $H^\infty = \lim_{n \to \infty} H[u_0(\cdot, t_n)]$, the following system of equations arises in the limit as $n \to \infty$:
\begin{align}
    |\partial\Omega_0^+| + |\partial\Omega_0^-| &= 1, \label{eq:pOm1}\\
    h^+(U_0^\infty) |\partial\Omega_0^+| + h^-(U_0^\infty) |\partial\Omega_0^-| &= U_0^\infty, \label{eq:hpmOmU}\\
    \Phi(h^+(U_0^\infty)) |\partial\Omega_0^+| + \Phi(h^-(U_0^\infty)) |\partial\Omega_0^-| + \frac{\alpha}{2} |U_0^\infty|^2 &= H^\infty. \label{eq:PhiOmH}
\end{align}

We now consider the above equations as an algebraic system with $U_0^\infty$ fixed and $|\partial \Omega_0^\pm|$ and $H^\infty$ as  variables. The conservation of area \eqref{eq:pOm1} ensures that $\frac{d}{d U_0^\infty}|\partial \Omega_0^+| = - \frac{d}{d U_0^\infty} |\partial \Omega_0^-|$, and differentiating \eqref{eq:hpmOmU} yields
\begin{equation}
   \frac{d h^+(U_0^\infty)}{d U_0^\infty}  |\partial\Omega_0^+| + \frac{d h^-(U_0^\infty)}{d U_0^\infty}  |\partial\Omega_0^-| + h^+(U_0^\infty) {d \over d U_0^\infty} |\partial\Omega_0^+| + h^-(U_0^\infty) {d \over d U_0^\infty} |\partial\Omega_0^-| = 1.
   \end{equation}
Rearranging the terms then establishes that $|\partial \Omega_0^+|$ is an increasing function of $U_0^\infty$, since 
\begin{equation}  
\label{eq:U0infOmp}
   {d \over d U_0^\infty} |\partial\Omega_0^+| =\frac{1 - |\partial\Omega_0^+| \frac{d h^+(U_0^\infty)}{d U_0^\infty} - |\partial\Omega_0^-| \frac{d h^-(U_0^\infty)}{d U_0^\infty}}{h^+(U_0^\infty) - h^-(U_0^\infty)} > 0.
\end{equation}

\begin{figure}
    \centering
    \includegraphics[scale = 0.5]{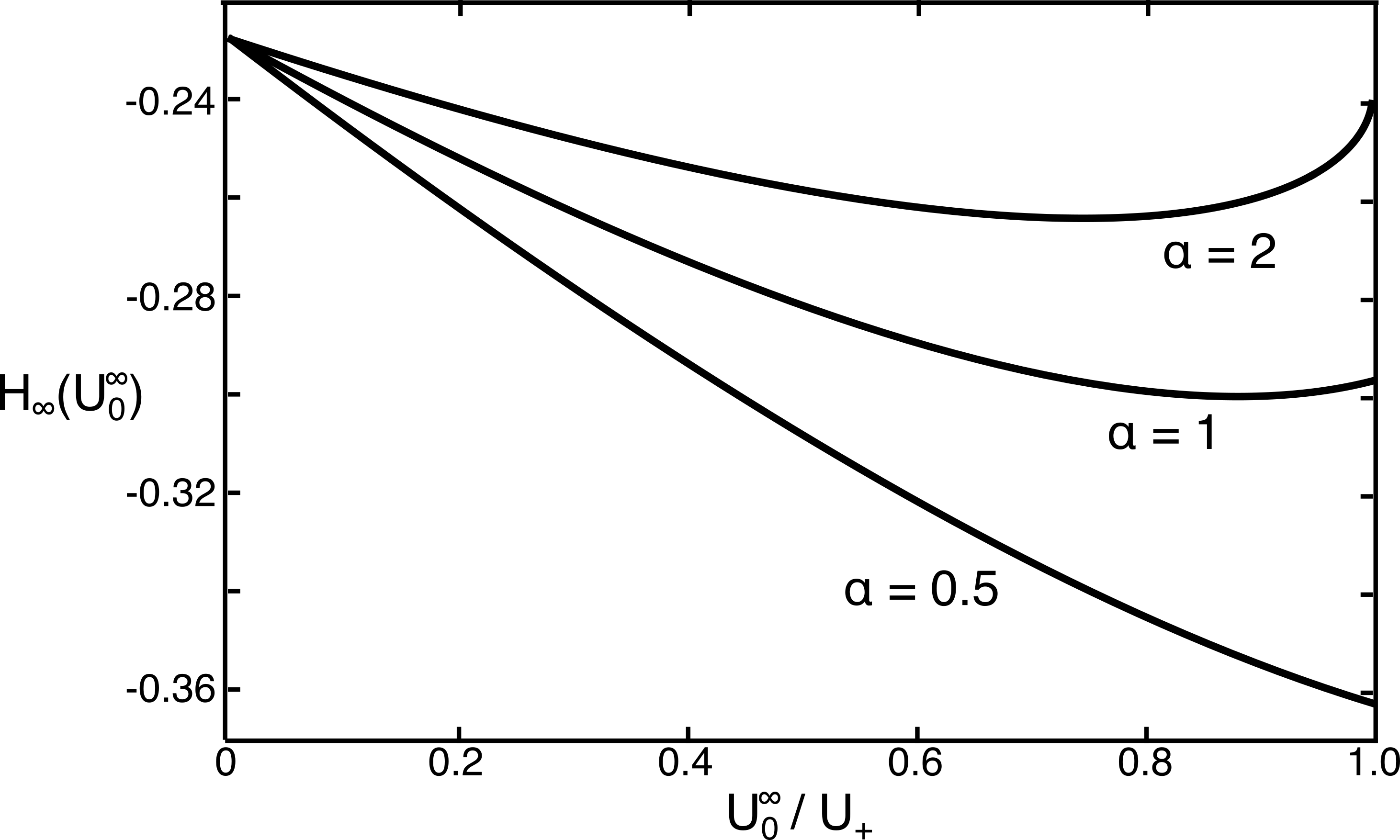}
    \caption{The function $H^\infty(U_0^\infty)$ solving \eqref{eq:pOm1}--\eqref{eq:PhiOmH}, plotted at different values of $\alpha$, with $\beta = 0.01$ and $\gamma = 0.25$.}
    \label{fig:conv}
\end{figure}

\noindent As the final step, taking the derivative of \eqref{eq:PhiOmH} and using the identity $\Phi'(h^\pm(U_0^\infty)) = -\alpha U_0^\infty$ yields the relationship
\begin{equation}
    {d H^\infty \over d U_0^\infty}  = \left(\frac{\Phi(h^+(U_0^\infty)) - \Phi(h^-(U_0^\infty)) }{h^+(U_0^\infty) - h^-(U_0^\infty)} - \Phi'(h^+(U_0^\infty))\right)(h^+(U_0^\infty) - h^-(U_0^\infty)) {d \over d U_0^\infty} |\partial\Omega_0^+|.
\end{equation}
The term inside the brackets only equals zero at a single value of $U_0^\infty$, corresponding to the double tangent construction (a straight line that touches the graph of $\Phi(u)$ at two points). For all other values of $U_0^\infty$, we thus have $dH^\infty / d U_0^\infty \ne 0$, as shown in Fig. \ref{fig:conv} for several choices of $\alpha$.  It follows that there are at most two isolated solutions $U_0^\infty$ to \eqref{eq:pOm1}--\eqref{eq:PhiOmH} for a given value of $H^\infty$, i.e., the set of possible values of $U_0^\infty$ for different choices of the sequence $(t_n)$ is discrete. Thus, since the $\omega$-limit set of $u_0$ is connected \cite{hale}, convergence of $H(u_0(\cdot, t_n)) \to H^\infty$ as $n \to \infty$ fixes $U_0^\infty$ independently of the subsequence. So we have $\lim_{t \to \infty} U_0(t) = U_0^\infty$, and by the ODE stability argument we then have $\lim_{t \to \infty} u(x, t) = h^\pm(U_0^\infty)$ as well for each $x \in \partial \Omega_0^\pm$, respectively. This arguments also shows that up to sets of measure zero the sets $\partial \Omega_0^\pm$ are independent of the sequence, and so we finally obtain a full limit: 
\begin{equation}
\label{eq:hpmlim}
    \lim_{t \to \infty} U_0(t) = U_0^\infty, \qquad \lim_{t \to \infty} u_0(x, t) = 
    \begin{cases}
    h^+(U_0^\infty) & x \in \partial \Omega_0^+, \\
    h^-(U_0^\infty) & x \in \partial \Omega_0^-,
    \end{cases}
\end{equation}
for some $\partial \Omega_0^\pm \subset \partial \Omega$ depending only on the parameters and the initial condition and satisfying $|\partial \Omega_0^-| + |\partial \Omega_0^+| = 1$. In other words, the solution of problem $(P_0^0)$ is expected to generically converge to a piecewise-constant function corresponding to the two stable equilibrium branches of the bistable nonlinearity for a certain long time limit value of $U_0$. This result can be interpreted as follows: for small but finite value of $\delta$ the solution of problem $(P_0^\delta)$ will become nearly piecewise constant on the timescale $1 \ll t \ll \delta^{-1}$ almost everywhere in $\partial \Omega$, except in a small transition region of width of order $\delta$. Thus, an interface is expected to form on this timescale, which will serve as an approximate initial condition for problem $(P_1^\delta)$ at the corresponding rescaled timescale $\delta \ll t \ll 1$ for the latter. 

 We note that while one should expect $u = h^\pm(U_0^\infty)$ to be generically selected as the limiting values of the solution away from the interfaces, for arbitrary initial conditions the possibility of $h^0(U_0^\infty)$ on a set of positive measure may not be \textit{a priori} excluded. This prevents us to make \eqref{eq:hpmlim} into a rigorous conclusion about the long-time limit behavior for problem $(P_0^0)$. This sort of issue is well known in the studies of gradient flows and its further treatment would require a more delicate analysis that goes beyond the scope of the present paper (for a treatment of a closely related problem, see \cite{ball2015quasistatic}).
 
Lastly, we point out that the arguments above do not \textit{a priori} exclude the possibility of $|\partial \Omega_0^+| = 0$ or $|\partial \Omega_0^-| = 0$, depending on the initial conditions. In these cases the system ends up in a uniform state and no further non-trivial dynamics governed by \eqref{eq:surfdiff} -- \eqref{eq:surfic}  is expected to occur. This outcome of the dynamics may be interpreted as a failure of cell polarization. However, such an outcome would be non-generic for the kinetic parameters corresponding to the regions $ii$ and $iii$ in Fig. \ref{fig:stab}, in which all uniform states are linearly unstable, consistently with the conclusion at the end of Sec. \ref{sec:pre}. In contrast, the case $0 < |\partial \Omega_0^+| < 1$ corresponds to a patterned state as the outcome of the dynamics of $(P_0^0)$, in which the boundary between $\partial \Omega_0^+$ and $\partial \Omega_0^-$ represents an interface between the high- and low-concentration domains, respectively.

\section{Convergence of surface concentration}\label{Sec:P2}

Having established the conditions for generation of the interface bounding uniform domains via non-local coupling, the next step is an examination of the evolution of these interfaces. At $O(\delta^{-1})$ timescale (in the original variables), the dynamics of our system is described by problem $(P_1^\delta)$. We claim that in the limit $\delta \to 0$ the solution $u_1^\delta$ of this problem converges to that of
\begin{equation}
    (P_1^0) \qquad
        \begin{aligned}
        u_1(x, t) & = 
        \begin{cases}
        h^+(U_1(t)) & x \in \partial \Omega_1^+(t), \\
        h^-(U_1(t)) & x \in \partial \Omega_1^-(t), 
        \end{cases}
        \qquad \partial \Omega_1^+(t) \cup \Gamma(t) \cup \partial \Omega_1^-(t) = \partial \Omega, \\
        U_1(t) & = h^+(U_1(t)) |\partial \Omega_1^+(t)| + h^-(U_1(t)) |\partial \Omega_1^-(t)|, \\
        {\partial \Gamma (x, t) \over \partial t} & = c(U_1(t)), \qquad x \in \Gamma(t). \label{eq:eik}
        \end{aligned}
\end{equation}
Here $\partial \Omega_1^\pm(t)$ are the time-dependent subsets of $\partial \Omega$ corresponding to the quasi-steady states $h^\pm(U_1(t))$ and $\Gamma(t) \subset \partial \Omega$ is a smooth closed curve or a collection of curves representing the boundary between $\partial \Omega_1^+(t)$ and $\partial \Omega_1^-(t)$, i.e., $\Gamma(t)$ is the interface. With some abuse of notation, we denote by $\partial \Gamma(x, t) / \partial t$ the normal velocity of $\Gamma(t)$ at each point  $x \in \Gamma(t)$ in the direction of $\partial \Omega_1^-(t)$, with the function $c(U_1)$ to be specified. As was already discussed at the end of Sec. \ref{sec:gen}, the sets $\partial \Omega_1^\pm(t)$ at $t = 0$ coincide with the sets $\partial \Omega_0^\pm$ obtained in the long time limit of the solution of problem $(P_0^0)$.

To aid in the derivation of $(P_1^0)$, it is convenient to define $\Psi = \Psi(x, t)$ which is the signed distance function from $\Gamma(t)$, with $\Psi > 0$ corresponding to $u_1 = h^+(U_1(t))$. Defining the stretched distance $z = \Psi(x, t) / \delta$ to $\Gamma(t)$, we seek the solution of $(P_1^\delta)$ in the neighborhood of $\Gamma(t)$ in the form $u_1^\delta(x, t) \simeq v(z, t)$, where $v$ is an unknown function \cite{fife}. Then to the leading order in $\delta$ we obtain in some tubular neighborhood of $\Gamma(t)$:
\begin{equation}
\label{eq:eikPsi}
    \partial_t \Psi(x, t) = c(U_1(t)), \qquad |\nabla \Psi(x, t)| = 1,
\end{equation}
where the function $c(U)$ is obtained by solving the traveling wave equation
\begin{equation}
\label{eq:TWU}
\partial^2_z v - c(U) \partial_z v - v + f(v) (1 - \alpha U) = 0, \qquad \lim_{z \to \pm \infty} v = h^\pm(U).  
\end{equation}
The interface $\Gamma(t)$ is then reconstructed from the condition $\Psi(x, t) = 0$ for $x \in \Gamma(t)$. It is well known that for all values of $U$ for which the nonlinearity in \eqref{eq:TWU} is of bistable type, this equation has a unique solution (up to translations) for a unique value of $c(U)$ \cite{fife77}. In particular, there is a unique value $U = U_1^\infty$ such that $c(U_1^\infty) = 0$, which is obtained by solving the equation 
\begin{equation}
\label{eq:maxwell}
    \int_{h^-(U)}^{h^+(U)} \left[(1 - \alpha U)f(u) - u \right] du = 0,
\end{equation}
and the sign of $c(U)$ coincides with the sign of the integral in \eqref{eq:maxwell} for $U \not= U_1^\infty$. It is also possible to show that $dc(U)/dU < 0$. Note that \eqref{eq:eikPsi} is equivalent to the last equation in \eqref{eq:eik}, with the function $c(U)$ given implicitly by the solution of \eqref{eq:TWU}. The rest of the equations in \eqref{eq:eik} are obtained by passing to the limit $\delta \to 0$ in the ansatz for $u_1^\delta$.

\begin{figure}[t]\label{Vinf}
    \centering
    \includegraphics[scale = 0.5]{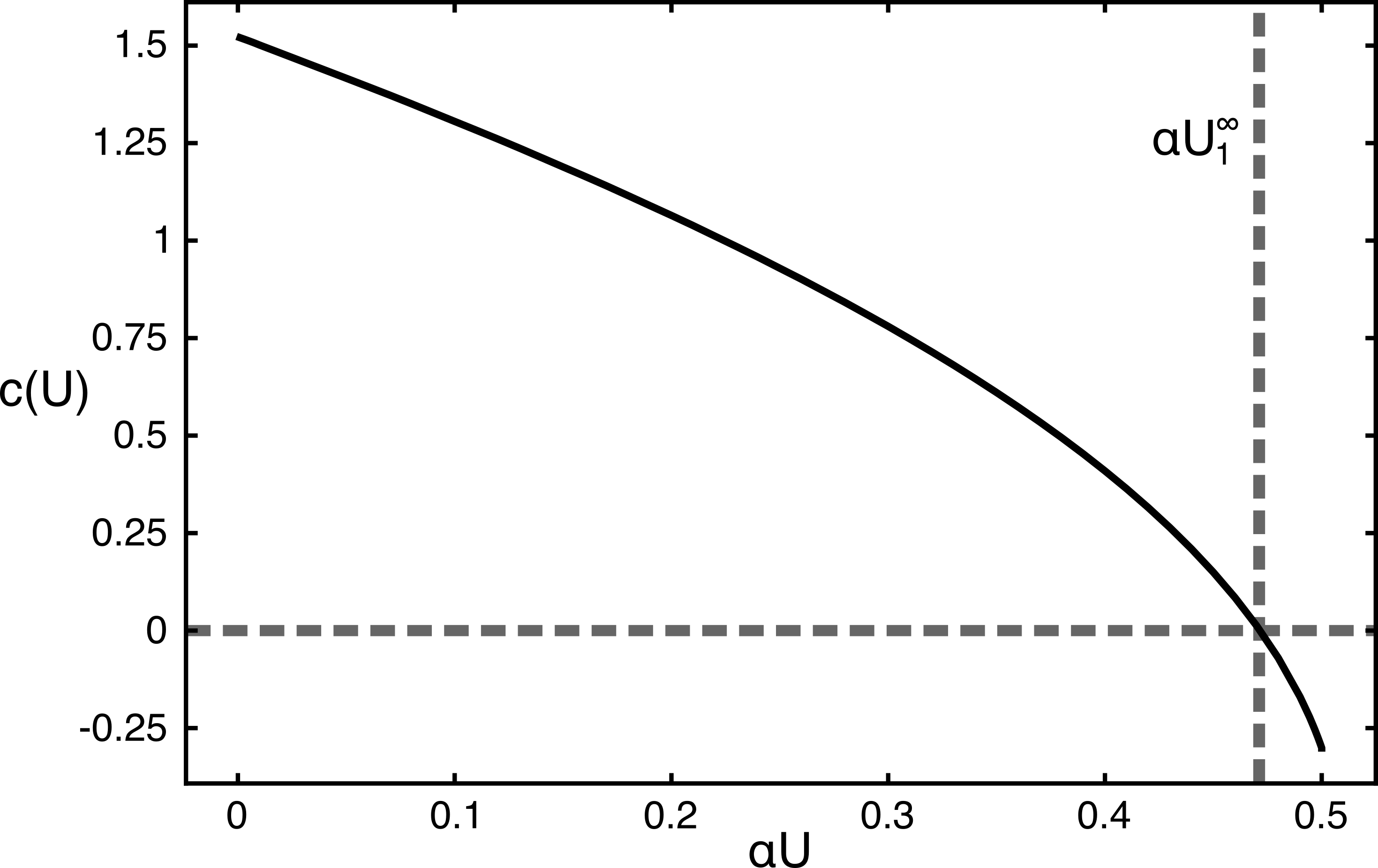}
    \caption{The dependence $c(U)$ obtained numerically for $\beta = 0.01$ and $\gamma = 0.25$. }
    \label{fig:cU}
\end{figure}

The dependence $c(U)$ obtained from the numerical solution of \eqref{eq:TWU} for a particular choice of the parameters is illustrated in Fig. \ref{fig:cU}. This calculation entailed time integration of 
\begin{align}
    \partial_t u(x, t) = \partial^2_{x} u(x, t) - u(x, t) + f(u(x, t)) (1-\alpha U)
\end{align} 
for fixed values of $U$, the initial condition
\begin{equation}
    u(x, 0) = \begin{cases}
    h^+(U) & x > 0, \\
    h^-(U) & x \le 0,
    \end{cases}
\end{equation}
and the boundary conditions  $\partial_x u(\pm L, 0) = 0$. 
Integration was performed using the Dedalus spectral package on a domain $[-L, L]$ with $L=100$, discretized into $1024$ modes \cite{2020PhRvR...2b3068B}. Each simulation was allowed to relax until $t=100$ with $dt = 0.1$, at which point the front velocity was calculated as 
\begin{align}
    c(U) = {\int_{-L}^L \partial_t u(x,t) dx \over h^+(U)-h^-(U)}.
\end{align} 

To study the long time behavior of the solutions of problem $(P_1^0)$, we argue as in \cite{henry2018multiple} and note that the growth rate of $|\partial \Omega_1^\pm(t)|$ is related to $c(U_1)$ as 
\begin{equation}
    {d \over dt} |\partial \Omega_+(t)| = c(U_1(t)) |\Gamma(t)|. 
\end{equation}
From this, we have
\begin{equation}
     {d U_1 \over dt}  = k(t) (U_1^\infty - U_1), 
\end{equation}
where
\begin{equation}
    \qquad k(t) = \left(\frac{ h^+(U_1(t)) - h^-(U_1(t))) }{1 -  {d h^{+}(U_1(t) \over dt} |\partial \Omega_1^+(t)| - {d h^{-}(U_1(t)) \over dt} |\partial \Omega_1^-(t)| } \right) {c(U_1(t)) |\Gamma(t)| \over U_1^\infty - U_1(t)}.
\end{equation}
As $0 < k(t) < \infty$ uniformly so long as the interface persists (recall that $dc(U) / dU < 0$), as $t \rightarrow \infty$ we find that $U_1(t) \to U_1^\infty$ exponentially fast and, therefore, $\Gamma(t) \to \Gamma_1^\infty$, with $\Gamma_1^\infty$ separating the two limiting domains $\partial \Omega_{1,\infty}^\pm$ in $\partial \Omega$ such that $\lim_{t \to \infty} u_1(x, t) = h^\pm(U_1^\infty)$ for all $x \in \partial \Omega_{1,\infty}^\pm$. Again, this behavior can be interpreted as follows:  for times $1 \ll t \ll \delta^{-1}$ the solution of problem $(P_1^\delta)$ will approach the long time limit of the solution of problem $(P_1^0)$, except in a transition layer of width $\delta$. This solution corresponds to the times $\delta \ll t \ll 1$ for problem $(P_2^\delta)$ and will, therefore, serve as the approximate initial condition for the latter.

An interesting feature of the long time behavior of problem $(P_1^0)$ is that as $t\rightarrow \infty$, the total amount $U$ of material on the surface converges to a value that depends only on $\alpha$, $\beta$, and $\gamma$, and is independent of the initial condition. Furthermore, the long time value of 
\begin{equation}
\label{eq:AOm2}
    |\partial \Omega_{1,\infty}^+| = {U_1^\infty - h^-(U_1^\infty) \over h^+(U_1^\infty) - h^-(U_1^\infty)}
\end{equation}
carries no explicit dependence of diffusion or domain geometry, other than through the dependence of the dimensionless parameter $\alpha$ on the surface to volume ratio $A / V$. The dependence of $U_1^\infty$ on $\alpha$ and $\gamma$ can be expressed as $U_1^\infty = \frac{1 - b(\beta) \gamma}{\alpha}$ for some function $b(\beta)$.   We illustrate this fact in Fig. \ref{Uinf}, which demonstrates the convergence of $U(t) \rightarrow U_\infty$ from three distinct initial conditions.

Some consideration is merited on the possible elimination of fronts: arbitrary initial conditions imply an arbitrary number of closed interfaces might be created during the initial dynamics, and some of these interfaces may vanish under the dynamics described above. However, provided we restrict ourselves to a parameter choice where $h^-(U_1^\infty(\alpha, \beta, \gamma)) < U_1^\infty(\alpha, \beta, \gamma) < h^+(U_1^\infty(\alpha, \beta, \gamma))$, as well as $U_- < U_1^\infty < U_+$ as per the previous section, both $h^+$ and $h^-$ will exist as real roots and conservation of mass will require $|\partial \Omega_1^+(t)|, \ |\partial \Omega_1^-(t)| > 0$ for all $t$. Thus, in this regime, at least one interface exists as $t \rightarrow \infty$.

\section{Area-preserving geodesic curvature flow}

In the previous section, we demonstrated that on the $O(\delta^{-1})$ timescale (in the original variables) the dynamics results in a stationary interface with a prescribed quantity of matter on the surface and a fixed area of the domains bounded by the interface. We now show that on longer timescales there is no longer a net transport of mass between the boundary $\partial \Omega$ and the enclosed volume $\Omega$, and  the dynamics may be reduced to a gradient flow driven by an interface length subject to domain area conservation \cite{mantegazza}. 

The dynamics on the $O(\delta^{-2})$ timescale (again, in the original variables) is described by problem $(P_2^\delta)$. We now claim that in the limit $\delta \to 0$ the solution $u_2^\delta$ of this problem converges to that of
\begin{equation}
    (P_2^0) \qquad
        \begin{aligned}
        u_2(x, t) & = 
        \begin{cases}
        h^+(U_1^\infty) & x \in \partial \Omega_2^+(t), \\
        h^-(U_1^\infty) & x \in \partial \Omega_2^-(t), 
        \end{cases}
        \qquad \partial \Omega_2^+(t) \cup \Gamma(t) \cup \partial \Omega_2^-(t) = \partial \Omega, \\
        U_1^\infty & = h^+(U_1^\infty) |\partial \Omega_2^+(t)| + h^-(U_1^\infty) |\partial \Omega_2^-(t)|, \\
        {\partial \Gamma (x, t) \over \partial t} & = k_g(x, t) + \xi(t), \qquad x \in \Gamma(t). \label{eq:mcf}
        \end{aligned}
\end{equation}
Here, as before, $\partial \Omega_2^\pm(t)$ are the time-dependent subsets of $\partial \Omega$ corresponding to the quasi-steady states $h^\pm(U_1^\infty)$ and $\Gamma(t) \subset \partial \Omega$ is a smooth closed curve or a collection of curves representing the boundary between $\partial \Omega_2^+(t)$ and $\partial \Omega_2^-(t)$. The function $k_g(x, t)$ refers to the geodesic curvature of  $\Gamma(t)$ at $x \in \Gamma(t)$, with the sign convention that $k_g < 0$ if $\partial \Omega_2^+$ is a small geodesic disk, and $\xi(t)$ is a Lagrange multiplier ensuring conservation of $|\partial \Omega_2^\pm(t)|$. The sets $\partial \Omega_2^\pm(t)$ at $t = 0$ now coincide with the sets $\partial \Omega_{1,\infty}^\pm$ obtained in the long time limit of the solution of problem $(P_1^0)$.

To derive this problem, we again denote by $z = \Psi(x, t)/\delta$ the stretched distance to the interface and seek the solution in the form \cite{fife}
\begin{equation}
    u_2^\delta(x, t) \simeq v( z - \zeta(t)) + \delta w ( z, x, t),
\end{equation}
where $v$, $w$ and $\zeta$ are to be found. Substituting this ansatz into $(P_2^\delta)$ yields
\begin{equation}
\begin{aligned}
    ( \partial_t \Psi & - \nabla^2 \Psi - \partial_t \zeta) \partial_z v  
    = \delta^{-1} \left[ \partial_{zz}^2 v - v + f(v)\left(1 -\alpha \int_{\partial \Omega} v dS \right) \right]   \\
    & + \partial_{zz}^2 w^\delta + \left[  - 1 + f'(v)\left(1 -\alpha \int_{\partial \Omega} v dS \right) \right]w -  \alpha f(v)\int_{\partial \Omega} w dS +  O(\delta). 
    \label{eq:vw}
\end{aligned}
\end{equation}
To eliminate the $O(\delta^{-1})$ term from the equation, we choose $v$ to be a solution of \eqref{eq:TWU} with $c(U) = 0$, which exists if and only if $\int_{\partial \Omega} v dS  = U_1^\infty$. By continuity, the latter can always be fixed by a suitable choice of $\zeta (t) = O(1)$ for a given $\Gamma(t)$ for which the condition $U_1^\infty = h^+(U_1^\infty) |\partial \Omega_2^+(t)| + h^-(U_1^\infty) |\partial \Omega_2^-(t)|$ holds.

Having now eliminated the $O(\delta^{-1})$ term from \eqref{eq:vw}, we can write the solvability condition for $w$, which is obtained by multiplying the equation by $\partial_z v$ and integrating over $z$ \cite{fife}:
\begin{equation}
    \left(\partial_t \zeta(t) - \partial_t \Psi(x,t) + \nabla^2 \Psi(y, t) \right) \int_{-\infty}^\infty |\partial_z v|^2 dz   - \alpha \int_{h^-(U_1^\infty)}^{h^+(U_1^\infty)} f(u) du \int_{\partial \Omega} w dS = O(\delta). \label{eq:sol}
\end{equation}
We note that if $y \in \Gamma(t)$ is the projection of $x \in \partial \Omega$ from a small tubular neighborhood of $\Gamma(t)$ onto $\Gamma(t)$, then $\nabla^2 \Psi(x, t) = \nabla^2 \Psi(y, t) + o(1)$ for $|x -y| \ll 1$. Thus, after dropping the $o(1)$ terms and rearranging the formula, we can rewrite \eqref{eq:sol} as
\begin{equation}
\partial_t \Psi(x, t) = \nabla^2 \Psi(y, t)  +  \xi(t), \qquad |\nabla \Psi(x, t)| = 1, \label{eq:fmcf}
\end{equation}
where 
\begin{equation}
\xi(t) =  \partial_t \zeta(t) - \frac{ \alpha \int_{h^-(U_1^\infty)}^{h^+(U_1^\infty)} f(u) du	}{\int_{-\infty}^{\infty} |\partial_z v|^2 dz} \int_{\partial \Omega} w(x, t) dS.
\end{equation}
The obtained equation \eqref{eq:fmcf} is nothing but the last equation in \eqref{eq:mcf} written in terms of the distance function \cite{ambrosio00a}, recalling the definition (with our sign convention) of geodesic curvature $k_g(x, t) = \nabla \cdot n(x, t)$, where $n(x, t) = \nabla \Psi(x, t)$ is the unit normal to the curve $\Gamma(t)$ at $x \in \Gamma(t)$ that lies in the tangent plane to $\partial \Omega$ at $x$ and points towards $\partial \Omega_2^+(t)$. Finally, conservation of $|\partial \Omega_2^\pm(t)|$ follows by passing to the limit $\delta \to 0$ in our ansatz, and the initial condition is given by the long time limit of the solution of problem $(P_1^0)$, i.e., we have $\Gamma(t) = \Gamma_1^\infty$ for $t = 0$. Note that the value of $|\partial \Omega_2^+|$ is given explicitly by the right-hand side of \eqref{eq:AOm2}.

Since the area of $|\partial \Omega_2^\pm(t)|$ is constant for the solutions of problem $(P_2^0)$, we have $\partial_t |\partial \Omega_2^\pm(t) | = 0$.
Furthermore, since the growth rate of a region's area is equal to the integral of its normal velocity, we can eliminate $\xi(t)$ to obtain a purely geometric expression for the normal velocity:   
\begin{equation}
{\partial \Gamma(x, t) \over \partial t} = k_g(x, t) - \langle k_g(\cdot, t) \rangle_{\Gamma(t)}, \qquad x \in \Gamma(t),
\end{equation}
where $\langle k_g(\cdot, t) \rangle_{\Gamma(t)} = \frac{1}{|\Gamma(t)| } \int_{\Gamma(t)} k_g(x, t) ds$ and $|\Gamma(t)|$ is the length of $\Gamma(t)$. 
This equation describes an area-preserving geodesic curvature flow on a surface. A geometric identity relates the evolution of interface length to curvature as $ \frac{d}{dt} |\Gamma(t)| = - \int_{\Gamma(t)} k_g(x, t) {\partial \Gamma(x, t) \over \partial t} ds$ \cite{kolavr2017area}. From this, we obtain an inequality 
\begin{equation}\label{varK}
\frac{1}{|\Gamma(t)|}\frac{d}{dt} |\Gamma(t)| = \langle k_g(\cdot, t) \rangle^2_{\Gamma(t)}- \langle k_g^2(\cdot, t) \rangle_{\Gamma(t)}  \le 0,
\end{equation}
which demonstrates that the interface length $|\Gamma(t)|$ is strictly decreasing except for when $k_g$ is constant along $\Gamma(t)$. The interpretation of this result is that for small $\delta$ the long time limits of the solutions of problem $(P_2^\delta)$  will be close to domains of prescribed area and locally minimal interfaces of constant curvature (geodesic disks). 

The problem of identifying locally minimal interfaces on a given surface has been studied at length over the years, and while no comprehensive solution exists, there is a considerable literature of results which can inform our understanding \cite{ros2001isoperimetric, morgan2000some, howards1999isoperimetric}. In particular, it is known that for every $0 < a < 1$ there exists a set $\partial \Omega^+ \subset \partial \Omega$ whose boundary $\Gamma$ globally minimizes the perimeter among all sets with  $|\partial \Omega^+| = a$ \cite{ros2001isoperimetric}. In this case $\Gamma$ is a smooth curve of constant geodesic curvature. Framed in terms of our reaction-diffusion problem, this means that for appropriate choices of $\alpha$, $\beta$, and $\gamma$, a non-uniform steady state for $(P_2^0)$ should always exist regardless of the particular shape of $\Omega$. Whether the solution of problem $(P_2^0)$ converges as $t \to \infty$ to a global minimizer, or, indeed, converges, however, is not known a priori. We note that due to the area conservation the interface $\Gamma(t)$ associated with the solutions of problem $(P_2^0)$ cannot vanish. Furthermore, due to the gradient flow nature of the dynamics the $\omega$-limit set of these solutions consists only of steady states, and convergence to one of the steady states is guaranteed if the set of the steady states is discrete. In particular, this would be the case if the global minimizer of the perimeter were the unique constant curvature solution with the prescribed area. We note that the latter would represent the most robust and reproducible scenario for cell polarization.

We can also use existing variational results to construct surfaces which yield some surprising solutions. It has been widely suggested that strictly mass-conserving reaction-diffusion equations will exhibit uninterrupted coarsening and admit steady states with only a single domain, but this phenomenon has only been studied in the Euclidean setting \cite{brauns2021wavelength, tateno2021interfacial}. Multi-domain polarization has in fact been observed in real cells, and failure to account for this is seen as a failure point of minimal models such as the one in this paper \cite{chiou2021cells}. However, we find that when geometric effects are properly considered, multi-cap solutions \textit{are} possible on some surfaces, which we now demonstrate by constructing examples. Per the theorems in Ref. \cite{ritore2001constant}, if $\partial \Omega$ is a surface of revolution about the $z$-axis that is symmetric with respect to reflections around the $xy$-plane, and if $\partial \Omega$  has Gaussian curvature $K(z)$ that is strictly increasing in $|z|$, then the locally minimal interfaces bounding a fixed area can be fully classified \cite{ritore2001constant}, and are of the types depicted in Fig. \ref{hourfig}. All stable interfaces lie in the planes $z = \text{const}$, which means $k_g$ can be written as a function of $z$, and it is convenient to define a function on $\partial \Omega$ of the form $L(z) = k_g^2(z) + K(z)$. In addition to single-interface solutions, there are two possible solutions  with disjoint interfaces. Case 1 features two interfaces at $\pm z_0$ for some $z_0 > 0$, respectively, which are stable provided they lie within the region where $L(z_0) < 0$. Case 2 has one interface at $z_1$ (s.t. $L(z_1) < 0)$ and the other at some $z_2$ such that $z_1 z_2 < 0$ and $L(z_2) > 0$. As the positions of $\partial \Omega^+$ and $\partial \Omega^-$ are interchangeable under geodesic curvature flow, each case can either represent a two cap solution, or a single high-concentration band containing $z = 0$.  In essence, these example solutions show the existence of a geometry-induced interruption to coarsening. As an aside, results such as those in Refs. \cite{ritore2001constant, ros2001isoperimetric, morgan2000some} do rule out multi-cap solutions on symmetric convex surfaces such as spheroids. 

\section{Numerical Study} \label{sec:num}

\begin{figure}[t]\label{hourfig}
    \centering
    \includegraphics[width = 0.9\textwidth]{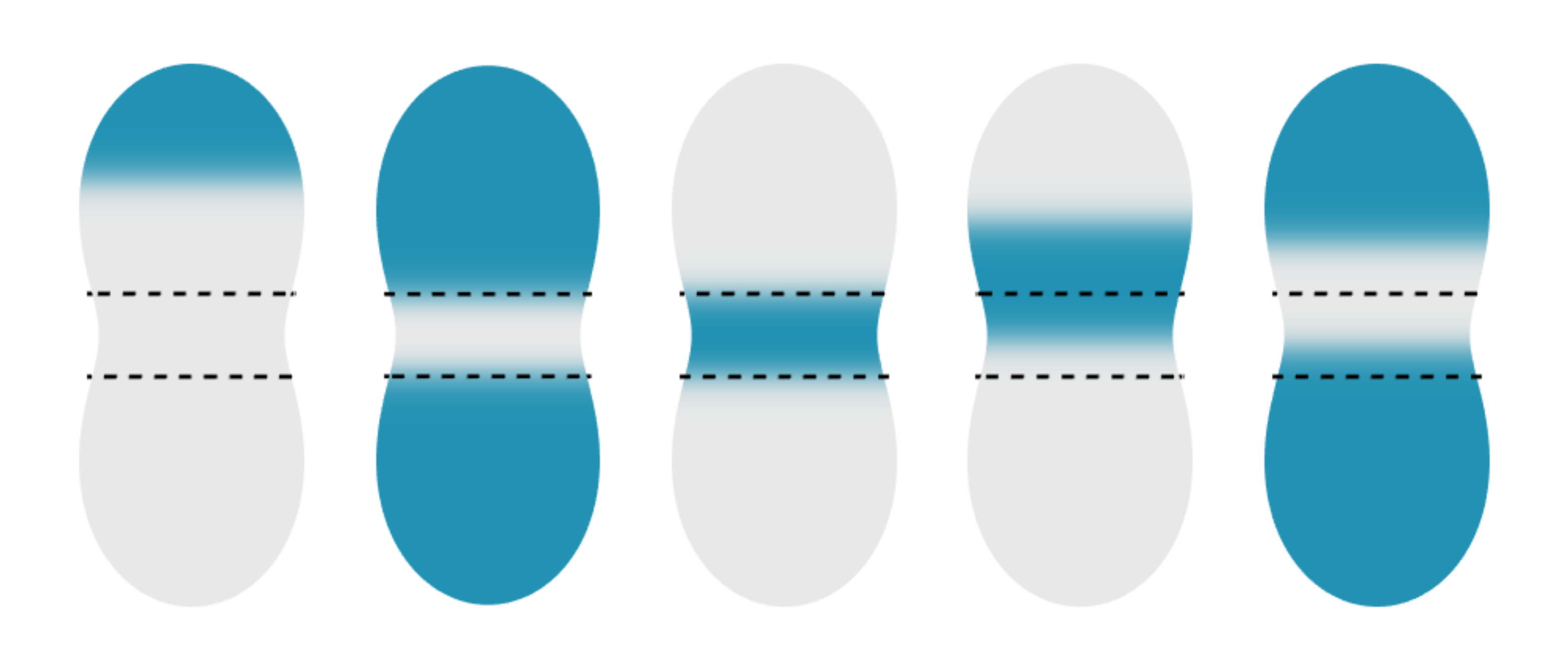}
    \caption{Numerical steady state solutions of \eqref{eq:surfdiff} -- \eqref{eq:surfic} on a surface of revolution (see the main text for details). The region of $L \le 0$, bounded by black dashed lines here, can support solutions with multiple stable interfaces.  Parameter choices are $\beta = 0.05$, $\gamma = 0.2$ and $\delta = 0.01$, and from left to right the values of $\alpha$ used were $10$, $3$, $13$, $10$ and $3.2$.  }
\end{figure}

We now verify numerically the existence of nontrivial steady state solutions predicted by the asymptotic analysis of $(P_2^\delta)$. We simulated \eqref{eq:surfdiff} -- \eqref{eq:surfic} on the surface shown in Fig. \ref{hourfig} for each of the possible configurations, using a recently developed numerical scheme \cite{miller2022forced}. The values of $\alpha$ used were (from left to right): $1$, $0.33$, $1.3$, $1$ and $0.35$. For all cases, $\beta = 0.05$, $\gamma = 0.2$ and $\delta = 0.01$. The surface geometry was constructed by revolution of polar curve $r(\theta) = 2 a^2 \left(\cos 2\theta + \sqrt{(b/a)^4 - \sin^2 2\theta}\right)$ (for $a = 0.2728$ and $b=0.3896$) about the vertical axis. The simulations were performed on a spatial discretization of $128 \times 128$ modes and integrated with $dt = 0.1$ until $t = 500$, at which point a numerical steady state was achieved. In each case, the initial conditions were step functions approximating the desired steady state. The steady states were verified using a GMRES-based Newton-Krylov method \cite{kelley2004newton}. This method further allowed us to compute approximate values of the dominant eigenvalues, and in each case these were negative, consistent with the proposed stability of these states. While the constraint on $U_\infty$ for the single interface case is as described in the previous section, both case 1 and 2 of the double interface solutions have a narrower range of acceptable $U_\infty$, owing to the constraints on the locations of the interfaces. As such, we chose a different $\alpha$ for each simulation to ensure a viable solution. 
 
\begin{figure}[t!]\label{Uinf}
    \centering
    \includegraphics[scale = 0.5]{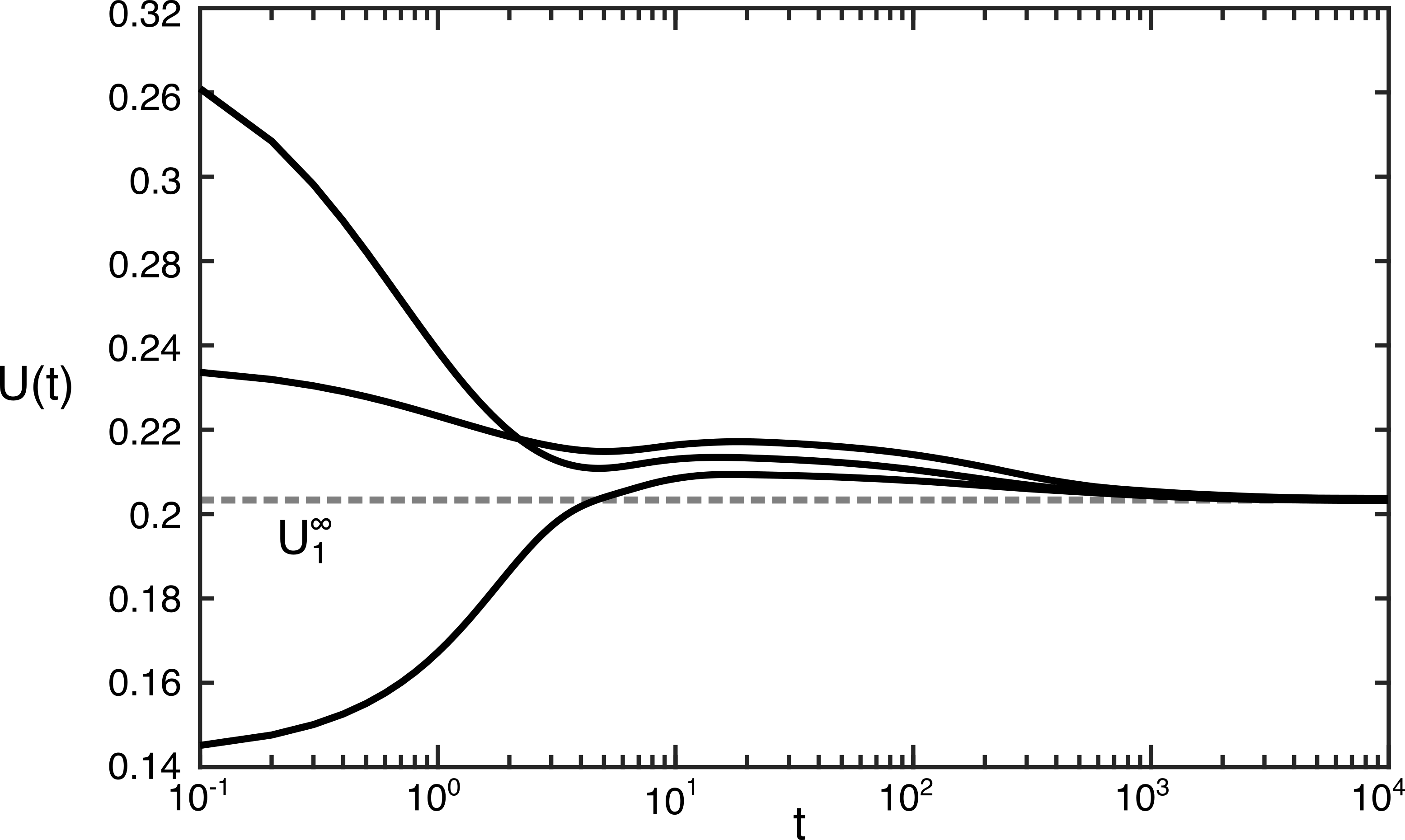}
    \caption{Numerical trajectories of $U(t)$ from three distinct initial conditions. The three simulations have different initial conditions, but the same model parameters, and converge to the value $U_1^\infty(\alpha, \beta, \gamma)$ from the asymptotic analysis of Sec. \ref{Sec:P2} as $t \rightarrow \infty$. Simulations performed at $\alpha = 2.25$, $\beta = 0.0001$, $\gamma = 0.25$ and $\delta = 10^{-3}$ on a sphere.}
\end{figure}

To corroborate the multiple stages of the dynamics predicted by our asymptotic analysis, we performed two additional numerical studies of \eqref{eq:surfdiff} -- \eqref{eq:surfic}, again using the integration methods introduced in Ref. \cite{miller2022forced}.  Recall that the  results of Sec. \ref{sec:gen} indicate convergence of $U(t)$ to a constant value $U_1$ depending on the initial condition on a large $O(1)$ time scale. Furthermore, by the results of Sec. \ref{Sec:P2} the function $U(t)$ converges to the final value $U_1^\infty(\alpha, \beta, \gamma)$ independent of initial conditions on a large $O(\delta^{-1})$ time scale. As the first test of this, we examine the convergence of $U(t)$ in three simulations. These were initialized at $u_0 + \gamma \sigma(x)/2$, where $u_0$ is as defined in Sec. \ref{sec:pre}, and $\sigma(x)$ is a random function taking on values between $-1$ and $1$, with the spatial correlation length $0.6$. Further details on how this distribution was generated can be found in the supplementary materials of Ref. \cite{miller2022forced}.  Simulations are performed on a sphere of unit area, discretized to $256 \times 256$ modes, and integrated with a timestep of $dt = 0.1$. The results of the simulations for $\delta = 10^{-3}$ are plotted in Fig. \ref{Uinf} (see the figure caption for the values of other parameters). The trajectories of all three examples can be split into two types of behavior. Dynamics from $t_0=0$ to $t_1\approx 10^1$ exhibit a rapid change in $U(t)$ towards an approximately constant value: this is the timescale of the interface generation. On the other hand, on the interval from $t_1 \approx 10^1$ through $t_2\approx 10^4$, the different instances of $U(t)$ converge to the predicted steady state at $U_1^\infty$ from Sec. \ref{Sec:P2}. 

\begin{figure}[t]
    \centering
    \includegraphics[scale = 0.5]{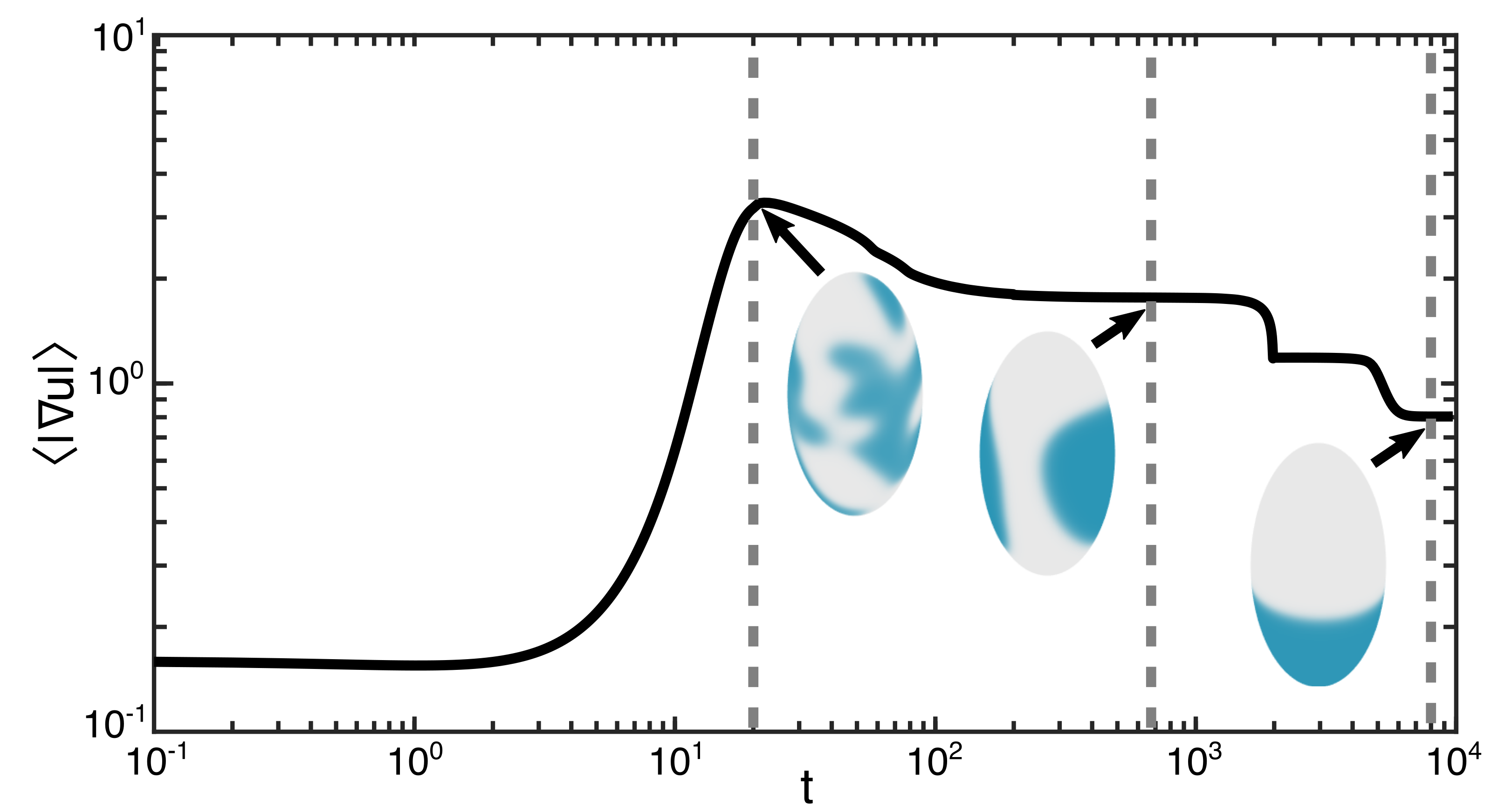}
    \caption{Tracking the  quantity $\langle |\nabla u| \rangle = \int_{\partial \Omega }|\nabla u| dS$ shows us the growth and shrinkage of the interface over time.  Evolution was simulated at $\alpha = 0.58, \beta = 0.001, \gamma = 0.4$, and $\delta = 0.01$. We observe differing dynamical regimes each roughly two decades in timescales: initial growth as the interface forms, followed by evolution under an eikonal-type flow, and further monotonic interface length decrease under geodesic curvature flow. The three inset plots show the simulation near the approximate end time of each regime. }
    \label{fig:surf_diff}
\end{figure}

 A second numerical experiment is to study the behavior of the interface length over time. For $\delta \ll 1$, an effective proxy for the total interface length is the quantity 
 \begin{equation}
     \langle | \nabla u | \rangle(t) = \int_{\partial \Omega} |\nabla u(\cdot, t) | dS,
 \end{equation}
 as in the last stage of the evolution the profile $u(x, t)$ will be close to two fixed values everywhere outside the thin transition layers approximating the interfaces. 
 We initialize a simulation at a small perturbation about the steady state as $u_0 + 0.01 \sigma(x)$ and plot the evolution of the interface in Fig. \ref{fig:surf_diff}. The simulation domain was a unit-area prolate spheroid with an aspect ratio of $2$, discretized into $128 \times 128$ modes. As before, our timestep was $dt = 0.1$. 
 
 For our choice of $\delta = 10^{-2}$ here, the distinct regimes corresponding to the limit behaviors of $(P_0^\delta)$, $(P_1^\delta)$ and $(P_2^\delta)$ as $\delta \to 0$ are very clearly expressed. Much as in the previous example, the timescale of interface formation is on the order of $t \approx 10^1$: the order of magnitude increase in $\delta$ between the two figures has a minimal effect on the zeroth order behavior of the system. At the end of this interval, surface concentration is fully partitioned into approximately uniform domains $\partial \Omega^\pm$. Once the quantity $\langle | \nabla u| \rangle$ peaks, it begins a period of slower evolution consistent with the predicted front motion of $(P_1^0)$, which then plateaus once $U_1^\infty$ is reached. A slower evolution of $\langle | \nabla u| \rangle$, which is now monotonically decreasing, subsequently occurs under the approximate $(P_2^0)$ dynamics. A brief note should be made of the step-like shape of the plot in this final interval. Contraction under $(P_1^0)$ front motion reduces this simulation to two spots, and the rapid drop in interface length which occurs around $t=2000$ represents the annihilation of one of these. After this, the remaining spot persists for a long period near midway between the two poles before finally migrating to one end of the spheroid. The final steady state is at the point of maximum Gaussian curvature, consistent with the geometric results noted in the previous section. Taken together, these numerical tests represent strong evidence that the formal asymptotics developed here accurately reproduces the behavior of  \eqref{eq:surfdiff} -- \eqref{eq:surfic}  for small but finite choices of $\delta$.

\section{Discussion} 

In summary, we performed an extensive analysis of the asymptotic behavior of a minimal model of cell polarization on a 2D manifold. Our results establish a hierarchy of three timescales of the dynamics in the case of small surface diffusivity. On the short timescale, we derived an evolution equation in the form of a globally coupled system of ODEs. While our initial problem itself appears to lack a variational structure, the reduced dynamics is shown to be a gradient flow generated by a nonlocal energy functional. From this, we were able to demonstrate generation of interfaces between high- and low-concentration regions for generic initial data. Our argument relied on the assumption that the solution for almost all points of the surface approached one of the two stable branches of the bistable nonlinearity, which is expected to be generically true but cannot be established for all initial data to make the result fully rigorous. Let us mention that in Ref. \cite{ball2015quasistatic}, in a closely related nonlocal Allen-Cahn problem, the authors were able to avoid this kind of assumption by constructing a second Lyapunov functional (in fact, an infinite family of Lyapunov functionals) to prove full convergence in time at the expense of allowing unstable solutions in the limit as well. However, the direct coupling between the nonlocal and nonlinear terms in problem $(P_0^0)$ greatly complicates this sort of argument, so it remains unclear whether this approach could be adapted to our problem. 

On the intermediate timescale, we demonstrated in the sharp-interface limit that the interfacial velocity is uniform and depends only on the enclosed area, and that this enclosed area approaches a final value determined only by the kinetic rates. This  area is determined strictly by the dimensionless parameters of our model, and thus is independent of the surface geometry. This fact raises a question as to how robust cell polarization can be to perturbations of the cell shape. This question could be particularly interesting in light of recent interest in coupled chemo-mechanical models of deforming cell membranes \cite{miller2018geometry, miller2020gait, elliott2021domain, gomez2021pattern}.

Finally, on the long timescale, we demonstrated that the dynamics reduce to area-preserving interface motion by geodesic curvature. This result formally establishes the intuitive notion that on curved surfaces wave-pinning phenomena is driven by a gradient flow generated by the interface length. The steady state of such a flow has a highly nontrivial dependence on surface shape - the ability of some surface geometries to support several distinct solutions suggests important implications for cellular decision-making and addresses a perceived failure of minimal mass-conserving reaction-diffusion models. One particularly interesting question for future work characterizing mechanisms to select between different steady states. Very recently theoretical work has shown that chemical gradients can enforce state transitions in simple geometries \cite{buttenschon2022cell}. Examining similar effects on geometries such as those in Fig. \ref{hourfig} could provide useful insight into biological control mechanisms. 

There is merit in some further comparison between our results and some deceptively similar results from flat surfaces. In the 1D version of this system, multi-domain solutions can also form, but these are strictly metastable and vanish on exponentially slow timescales \cite{mori2011asymptotic, sun2000dynamics}. By contrast, the multi-domain solutions discussed here can be truly stable. Previous works such as Ref. \cite{vanderlei2011computational} have also showed that domain geometry can establish multi-domain solutions in wave-pinning models. One key difference is that these studies examined flat 2D domains bounded by no-flux boundaries of various shapes, rather than curved surfaces. Both versions can produce multi-domain solutions, but a key difference lies in the time needed to reach steady states. Specifically, a circular domain in the flat scenario will be metastable if it is sufficiently far from the boundary, and like in the 1D case discussed above it will only approach the steady state location exponentially slowly. This slowness stems from the approximate translational invariance of the interfaces away from the boundaries. Non-uniform local curvature breaks this invariance, making the convergence to steady state domain locations considerably faster, on the algebraic rather than exponentially long timescale in terms of the inverse diffusivity. 

The wave-pinning model studied here represents one of the simplest possible characterizations of coupled bulk-surface dynamics in a single cell. A great number of more complex models have been proposed over the past decade, including ones featuring multiple chemical species \cite{paquin2019pattern} and advection \cite{gross2019guiding, barberi2022localized}. Additionally, recent experiments have demonstrated that bulk-surface systems are capable of much more complex pattern formation than discussed here, such as the spiral wave dynamics observed in starfish embryo \cite{liu2021topological, tan2020topological}, establishing a long-term need for further analysis of this class of system. 

\bibliographystyle{siamplain}

\begin{thebibliography}{10}

\bibitem{alfaro12}
{\sc M.~Alfaro and H.~Matano}, {\em On the validity of formal asymptotic
  expansions in {A}llen-{C}ahn equation and {F}itz{H}ugh-{N}agumo system with
  generic initial data}, Discrete Contin. Dyn. Syst. Ser. B, 17 (2012),
  pp.~1639--1649.

\bibitem{ambrosio00a}
{\sc L.~Ambrosio}, {\em Geometric evolution problems, distance function and
  viscosity solutions}, in Calculus of Variations and Partial Differential
  Equations: Topics on Geometrical Evolution Problems and Degree Theory,
  G.~Buttazzo, A.~Marino, and M.~Murthy, eds., Springer, Berlin, Heidelberg,
  2000, pp.~5--93.

\bibitem{ball2015quasistatic}
{\sc J.~M. Ball and Y.~{\c{S}}eng{\"u}l}, {\em Quasistatic nonlinear
  viscoelasticity and gradient flows}, J. Dyn. Differ. Equ., 27 (2015),
  pp.~405--442.

\bibitem{barberi2022localized}
{\sc L.~Barberi and K.~Kruse}, {\em Localized states in active fluids},
  arXiv:2209.02581,  (2022).

\bibitem{bellettini95}
{\sc G.~Bellettini and M.~Paolini}, {\em Quasi-optimal error estimates for the
  mean curvature flow with a forcing term}, Differ. Integral Equ., 8 (1995),
  pp.~735 -- 752.

\bibitem{bialecki2017polarization}
{\sc S.~Bialecki, B.~Kazmierczak, and T.~Lipniacki}, {\em Polarization of
  concave domains by traveling wave pinning}, PloS one, 12 (2017), p.~e0190372.

\bibitem{brauns2021wavelength}
{\sc F.~Brauns, H.~Weyer, J.~Halatek, J.~Yoon, and E.~Frey}, {\em Wavelength
  selection by interrupted coarsening in reaction-diffusion systems}, Phys.
  Rev. Lett., 126 (2021), p.~104101.

\bibitem{2020PhRvR...2b3068B}
{\sc K.~J. {Burns}, G.~M. {Vasil}, J.~S. {Oishi}, D.~{Lecoanet}, and B.~P.
  {Brown}}, {\em {Dedalus: A flexible framework for numerical simulations with
  spectral methods}}, Phys. Rev. Research, 2 (2020), p.~023068.

\bibitem{buttenschon2022cell}
{\sc A.~Buttensch{\"o}n and L.~Edelstein-Keshet}, {\em Cell repolarization: A
  bifurcation study of spatio-temporal perturbations of polar cells}, Bull.
  Math. Biol., 84 (2022), pp.~1--29.

\bibitem{cartan}
{\sc H.~Cartan}, {\em Calcul Differentiel: {I-Calcul} differetiel dans les
  espaces de {Banach}; {II-Equations} differentielles}, Cours de mathematiques
  II, Hermann et Cie, Editeurs, 1967.

\bibitem{chen92a}
{\sc X.~Chen}, {\em Generation and propagation of interfaces for
  reaction-diffusion equations}, J. Differ. Equ., 96 (1992), pp.~116--141.

\bibitem{chen92}
{\sc X.~Chen}, {\em Generation and propagation of interfaces in
  reaction-diffusion systems}, Trans. Amer. Math. Soc., 334 (1992),
  pp.~877--913.

\bibitem{chiou2021cells}
{\sc J.-G. Chiou, K.~D. Moran, and D.~J. Lew}, {\em How cells determine the
  number of polarity sites}, eLife, 10 (2021), p.~e58768.

\bibitem{cusseddu2019coupled}
{\sc D.~Cusseddu, L.~Edelstein-Keshet, J.~A. Mackenzie, S.~Portet, and
  A.~Madzvamuse}, {\em A coupled bulk-surface model for cell polarisation}, J.
  Theor. Biol., 481 (2019), pp.~119--135.

\bibitem{diegmiller2018spherical}
{\sc R.~Diegmiller, H.~Montanelli, C.~B. Muratov, and S.~Y. Shvartsman}, {\em
  Spherical caps in cell polarization}, Biophys. J., 115 (2018), pp.~26--30.

\bibitem{edelstein2013simple}
{\sc L.~Edelstein-Keshet, W.~R. Holmes, M.~Zajac, and M.~Dutot}, {\em From
  simple to detailed models for cell polarization}, Philos. Trans. R. Soc.
  Lond., B, 368 (2013), p.~20130003.

\bibitem{elliott2021domain}
{\sc C.~M. Elliott and L.~Hatcher}, {\em Domain formation via phase separation
  for spherical biomembranes with small deformations}, Eur. J. Appl. Math., 32
  (2021), pp.~1127--1152.

\bibitem{fife}
{\sc P.~C. Fife}, {\em Dynamics of Internal Layers and Diffusive Interfaces},
  Society for Industrial and Applied Mathematics, Philadelphia, 1988.

\bibitem{fife77}
{\sc P.~C. Fife and J.~B. McLeod}, {\em The approach of solutions of nonlinear
  diffusion equations to traveling front solutions}, Arch. Rational Mech.
  Anal., 65 (1977), pp.~335--361.

\bibitem{gessele2020geometric}
{\sc R.~Ge{\ss}ele, J.~Halatek, L.~W{\"u}rthner, and E.~Frey}, {\em Geometric
  cues stabilise long-axis polarisation of {PAR} protein patterns in \emph{{C}.
  elegans}}, Nat. Commun., 11 (2020), pp.~1--12.

\bibitem{ghose2022orientation}
{\sc D.~Ghose, T.~Elston, and D.~Lew}, {\em Orientation of cell polarity by
  chemical gradients}, Annu. Rev. Biophys., 51 (2022), pp.~431--451.

\bibitem{gomez2021pattern}
{\sc D.~Gomez, S.~Iyaniwura, F.~Paquin-Lefebvre, and M.~Ward}, {\em Pattern
  forming systems coupling linear bulk diffusion to dynamically active
  membranes or cells}, Philos. Trans. R. Soc. A, 379 (2021), p.~20200276.

\bibitem{gross2019guiding}
{\sc P.~Gross, K.~V. Kumar, N.~W. Goehring, J.~S. Bois, C.~Hoege,
  F.~J{\"u}licher, and S.~W. Grill}, {\em Guiding self-organized pattern
  formation in cell polarity establishment}, Nat. Phys., 15 (2019),
  pp.~293--300.

\bibitem{halatek2018rethinking}
{\sc J.~Halatek and E.~Frey}, {\em Rethinking pattern formation in
  reaction--diffusion systems}, Nat. Phys., 14 (2018), pp.~507--514.

\bibitem{hale}
{\sc J.~K. Hale}, {\em Asymptotic behavior of dissipative systems}, vol.~25 of
  Mathematical Surveys and Monographs, American Mathematical Society,
  Providence, RI, 1988.

\bibitem{hausberg2018well}
{\sc S.~Hausberg and M.~R{\"o}ger}, {\em Well-posedness and fast-diffusion
  limit for a bulk--surface reaction--diffusion system}, Nonlinear Differ. Equ.
  Appl., 25 (2018), pp.~1--32.

\bibitem{henry2018multiple}
{\sc M.~Henry, D.~Hilhorst, and C.~B. Muratov}, {\em A multiple scale pattern
  formation cascade in reaction-diffusion systems of activator-inhibitor type},
  Interfaces Free Boundaries, 20 (2018), pp.~297--336.

\bibitem{howards1999isoperimetric}
{\sc H.~Howards, M.~Hutchings, and F.~Morgan}, {\em The isoperimetric problem
  on surfaces}, Am. Math. Mon., 106 (1999), pp.~430--439.

\bibitem{kelley2004newton}
{\sc C.~T. Kelley, I.~Kevrekidis, and L.~Qiao}, {\em Newton-krylov solvers for
  time-steppers}, arXiv:math/0404374,  (2004).

\bibitem{kolavr2017area}
{\sc M.~Kol{\'A}{\v{R}}, M.~Bene{\v{s}}, and D.~{\v{S}}ev{\v{C}}ovi{\v{C}}},
  {\em Area preserving geodesic curvature driven flow of closed curves on a
  surface}, Discrete Contin. Dyn. Syst. - B., 22 (2017), pp.~3671--3689.

\bibitem{li2021bulk}
{\sc J.~Li, L.~Su, X.~Wang, and Y.~Wang}, {\em Bulk-surface coupling:
  Derivation of two models}, J. Differ. Equ., 289 (2021), pp.~1--34.

\bibitem{liu2021topological}
{\sc J.~Liu, J.~F. Totz, P.~W. Miller, A.~D. Hastewell, Y.-C. Chao, J.~Dunkel,
  and N.~Fakhri}, {\em Topological braiding and virtual particles on the cell
  membrane}, Proc. Natl. Acad. Sci. U.S.A., 118 (2021), p.~e2104191118.

\bibitem{madzvamuse2015stability}
{\sc A.~Madzvamuse, A.~H. Chung, and C.~Venkataraman}, {\em Stability analysis
  and simulations of coupled bulk-surface reaction--diffusion systems}, Proc.
  R. Soc. A, 471 (2015), p.~20140546.

\bibitem{mantegazza}
{\sc C.~Mantegazza}, {\em Lecture Notes on Mean Curvature Flow}, vol.~290 of
  Progress in Mathematics, Birkhauser, Basel, 2011.

\bibitem{miller2020gait}
{\sc P.~W. Miller and J.~Dunkel}, {\em Gait-optimized locomotion of wave-driven
  soft sheets}, Soft Matter, 16 (2020), pp.~3991--3999.

\bibitem{miller2022forced}
{\sc P.~W. Miller, D.~Fortunato, C.~Muratov, L.~Greengard, and S.~Shvartsman},
  {\em Forced and spontaneous symmetry breaking in cell polarization}, Nat.
  Comput. Sci., 2 (2022), pp.~504--511.

\bibitem{miller2018geometry}
{\sc P.~W. Miller, N.~Stoop, and J.~Dunkel}, {\em Geometry of wave propagation
  on active deformable surfaces}, Phys. Rev. Lett., 120 (2018), p.~268001.

\bibitem{morgan2000some}
{\sc F.~Morgan and D.~L. Johnson}, {\em Some sharp isoperimetric theorems for
  {R}iemannian manifolds}, Indiana Univ. Math. J.,  (2000), pp.~1017--1041.

\bibitem{mori2008wave}
{\sc Y.~Mori, A.~Jilkine, and L.~Edelstein-Keshet}, {\em Wave-pinning and cell
  polarity from a bistable reaction-diffusion system}, Biophys. J., 94 (2008),
  pp.~3684--3697.

\bibitem{mori2011asymptotic}
{\sc Y.~Mori, A.~Jilkine, and L.~Edelstein-Keshet}, {\em Asymptotic and
  bifurcation analysis of wave-pinning in a reaction-diffusion model for cell
  polarization}, SIAM J. Appl. Math., 71 (2011), pp.~1401--1427.

\bibitem{niethammer2020bulk}
{\sc B.~Niethammer, M.~R{\"o}ger, and J.~J. Vel{\'a}zquez}, {\em A bulk-surface
  reaction-diffusion system for cell polarization}, Interfaces Free Boundaries,
  22 (2020), pp.~85--117.

\bibitem{otsuji2007mass}
{\sc M.~Otsuji, S.~Ishihara, C.~Co, K.~Kaibuchi, A.~Mochizuki, and S.~Kuroda},
  {\em A mass conserved reaction--diffusion system captures properties of cell
  polarity}, PloS Comput. Biol., 3 (2007), p.~e108.

\bibitem{paquin2019pattern}
{\sc F.~Paquin-Lefebvre, W.~Nagata, and M.~J. Ward}, {\em Pattern formation and
  oscillatory dynamics in a two-dimensional coupled bulk-surface
  reaction-diffusion system}, SIAM J. Appl. Dyn., 18 (2019), pp.~1334--1390.

\bibitem{pisante2015allen}
{\sc A.~Pisante and F.~Punzo}, {\em {Allen--Cahn} approximation of mean
  curvature flow in riemannian manifolds, {II}: Brakke's flows}, Commun.
  Contemp. Math., 17 (2015), p.~1450041.

\bibitem{pisante2013allen}
{\sc A.~Pisante and F.~Punzo}, {\em {Allen-Cahn} approximation of mean
  curvature flow in riemannian manifolds {I}, uniform estimates}, Ann. Sc.
  Norm. Super. Pisa Cl. Sci. (5), 15 (2016), pp.~309--341.

\bibitem{rappel2017mechanisms}
{\sc W.-J. Rappel and L.~Edelstein-Keshet}, {\em Mechanisms of cell
  polarization}, Curr. Opin. Syst. Biol., 3 (2017), pp.~43--53.

\bibitem{ratz2015turing}
{\sc A.~R{\"a}tz}, {\em Turing-type instabilities in bulk--surface
  reaction--diffusion systems}, J. Comput. Appl. Math., 289 (2015),
  pp.~142--152.

\bibitem{ratz2014symmetry}
{\sc A.~R{\"a}tz and M.~R{\"o}ger}, {\em Symmetry breaking in a bulk--surface
  reaction--diffusion model for signalling networks}, Nonlinearity, 27 (2014),
  pp.~1805--1827.

\bibitem{ritore2001constant}
{\sc M.~Ritor{\'e}}, {\em Constant geodesic curvature curves and isoperimetric
  domains in rotationally symmetric surfaces}, Commun. Anal. Geom., 9 (2001),
  pp.~1093--1138.

\bibitem{ros2001isoperimetric}
{\sc A.~Ros}, {\em The isoperimetric problem}, in Global theory of minimal
  surfaces, D.~A. Hoffman, ed., vol.~2 of Clay mathematics proceedings,
  American Mathematical Society, Providence, Rhode Island, 2001, pp.~175--209.

\bibitem{sakamoto2000spatial}
{\sc K.~Sakamoto}, {\em Spatial homogenization and internal layers in a
  reaction-diffusion system}, Hiroshima Math. J., 30 (2000), pp.~377--402.

\bibitem{sharma2016global}
{\sc V.~Sharma and J.~Morgan}, {\em Global existence of solutions to
  reaction-diffusion systems with mass transport type boundary conditions},
  SIAM J. Math. Anal., 48 (2016), pp.~4202--4240.

\bibitem{sun2000dynamics}
{\sc X.~Sun and M.~J. Ward}, {\em Dynamics and coarsening of interfaces for the
  viscous cahn—hilliard equation in one spatial dimension}, Stud. Appl.
  Math., 105 (2000), pp.~203--234.

\bibitem{tan2020topological}
{\sc T.~H. Tan, J.~Liu, P.~W. Miller, M.~Tekant, J.~Dunkel, and N.~Fakhri},
  {\em Topological turbulence in the membrane of a living cell}, Nat. Phys., 16
  (2020), pp.~657--662.

\bibitem{tateno2021interfacial}
{\sc M.~Tateno and S.~Ishihara}, {\em Interfacial-curvature-driven coarsening
  in mass-conserved reaction-diffusion systems}, Phys. Rev. Research, 3 (2021),
  p.~023198.

\bibitem{trong2014parameter}
{\sc P.~K. Trong, E.~M. Nicola, N.~W. Goehring, K.~V. Kumar, and S.~W. Grill},
  {\em Parameter-space topology of models for cell polarity}, New J. Phys., 16
  (2014), p.~065009.

\bibitem{vanderlei2011computational}
{\sc B.~Vanderlei, J.~J. Feng, and L.~Edelstein-Keshet}, {\em A computational
  model of cell polarization and motility coupling mechanics and biochemistry},
  Multiscale Model. Simul., 9 (2011), pp.~1420--1443.

\bibitem{zhu2020developmental}
{\sc M.~Zhu, J.~Cornwall-Scoones, P.~Wang, C.~E. Handford, J.~Na, M.~Thomson,
  and M.~Zernicka-Goetz}, {\em Developmental clock and mechanism of de novo
  polarization of the mouse embryo}, Science, 370 (2020), p.~eabd2703.

\end{thebibliography}

\end{document}